\documentclass[11pt]{article}
\pdfoutput=1 

\usepackage[T1]{fontenc}
\usepackage[utf8]{inputenc}						
\usepackage{lmodern}%uses lmodern font as default throughout document								
\usepackage{textcomp}%adds additional glyphs to lmodern font											
\usepackage{microtype}%advanced typesetting (makes everything a bit more pretty)						
\usepackage[inline]{enumitem}%adds options for list environments										
\usepackage{scalerel}%scales glyphs such as \parallel
\usepackage{subdepth}%places subscripts a bit lower if there is also a superscript
\usepackage{ulem}
\usepackage{geometry}
\usepackage{subfiles}
\usepackage{import}
\usepackage{appendix}%adds subappendices

\usepackage[british]{babel}%fixes hyphenation in english, adds additional commands such as \"o for ö	
\usepackage[style=english]{csquotes}%required by babel, fixes quotation style 							
\usepackage{alphabeta}%adds greek letters outside of mathmode											
\usepackage{upgreek}
\usepackage[all,british]{foreign}%adds commands for abreviations such as e.g. and i.e.				
\usepackage{xcolor}
\usepackage{verbatim}

\usepackage[subrefformat = parens]{subcaption}%adds subfigure and subtable environments					
\usepackage{booktabs}%adds options for professional looking tables (look at documentation!)	
\usepackage{multirow}%adds option for centered input for multirows										
\usepackage{graphicx}
\usepackage{tabularx}							
\usepackage{makecell}							
\usepackage[export = true]{adjustbox}

\usepackage{mathtools}
\usepackage{amssymb}
\usepackage{dsfont}
\usepackage{physics}						
\usepackage{slashed}						
\usepackage{units}
\usepackage{bm}%fixes mathbf environment, use \bm command instead of \mathbf

\allowdisplaybreaks[1]%allows page breaks in equations but avoids them as much as possible

\let\mathbf\bm
\makeatletter
\g@addto@macro\bfseries{\boldmath}
\makeatother	

\usepackage{tikz}
\usepackage{tikzscale}
\usepackage{tikz-feynman}

\usepackage[affil-it]{authblk}%adds options for multiple authors and affiliations						

\usepackage{orcidlink}%insert link to orcid page
\usepackage[sorting = none]{biblatex}
\usepackage{bibentry}
\usepackage[nottoc]{tocbibind}
\usepackage{hyperref}
\hypersetup{
  hidelinks,
  colorlinks=true,
  linkcolor={black},
  citecolor={blue!50!black},
  linktoc=all,
  pdfencoding=auto,
  psdextra,
  breaklinks=true,
  pdfcreator={},
  pdfproducer={}
}
\usepackage[capitalize,nameinlink]{cleveref}%helps with crossfererencing, e.g. write "\cref{sec:...}" instead of "Section \ref{sec:...}"

\usepackage{hep-acronym}

\acronym{QFT}{quantum field theory}[quantum field theories]
\acronym{QCD}{quantum chromodynamics}
\acronym{QED}{quantum electrodynamics}
\acronym{GD}{gluon dynamics}
\acronym{SM}{Standard Model}
\acronym{RGE}{renormalization group equation}
\acronym{VEV}{vacuum expectation value}
\acronym{CKM}{Cabibbo-\allowbreak Kobayashi-\allowbreak Maskawa}
\acronym{PMNS}{Pontecorvo–\allowbreak Maki–\allowbreak Nakagawa–\allowbreak Sakata}
\acronym[$\chi$PT]{cPT}{chiral perturbation theory}[chiral perturbation theories]
\acronym{NGB}{Nambu-\allowbreak Goldstone boson}
\acronym{PNGB}{pseudo Nambu-\allowbreak Goldstone boson}
\acronym{WZW}{Wess-\allowbreak Zumino-\allowbreak Witten}
\acronym{MC}{Maurer-\allowbreak Cartan}
\acronym{GM}{Gell-\allowbreak Mann}
\acronym{CS}{Chern-\allowbreak Simons}
\acronym{YM}{Yang-\allowbreak Mills}
\acronym{HNL}{heavy neutral lepton}
\acronym{ALP}{axion-like particle}
\acronym{LLP}{long-lived particle}
\acronym{EOM}{equation of motion}
\acronym{PI}{partial integration}
\acronym[1PI]{onePI}{one particle irreducible}
\acronym{BSM}{beyond the Standard Model}
\longacronym{WIMP}{weakly interacting massive particle}
\acronym{DM}{dark matter}
\acronym{UV}{ultraviolet}
\acronym{IR}{infrared}
\acronym{LEC}{low energy coefficient}
\acronym{LER}{low energy realisation}
\acronym{NP}{new physics}
\acronym{EM}{electromagnetic}
\acronym{EW}{electroweak}
\acronym{EWSB}{electroweak symmetry breaking}
\acronym[\overline{MS}]{MSb}{modified minimal subtraction}
\acronym{NDA}{naive dimensional analysis}
\acronym{DOF}{degree of freedom}[degrees of freedom]
\acronym{LO}{leading order}
\acronym{NLO}{next-to-leading order}
\acronym{NNLO}{next-to-next-to-leading order}
\acronym{LHC}{Large Hadron Collider}
\shortacronym{CMS}{compact muon solenoid}
\shortacronym{ATLAS}{a toroidal LHC apparatus}
\shortacronym{LHCb}{LHC beauty}
\shortacronym{NA}*{North Area experiment \textnumero~}
\shortacronym{KOTO}{K0 at Tokai}
\shortacronym{SHiP}{Search for Hidden Particles}
\shortacronym{MATHUSLA}{Massive Timing Hodoscope for Ultra-Stable neutraL pArticles}
\shortacronym{FASER}{ForwArd Search ExpeRiment}
\shortacronym[CODEX-b]{CODEXb}{compact detector for exotics at LHCb}
\shortacronym{CP}{charge-parity}
\acronym{PET}{\emph{portal effective theory}}[\emph{portal effective theories}]
\acronym{EFT}{effective field theory}[effective field theories]
\acronym{SMEFT}{Standard Model effective field theory}[Standard Model effective field theories]
\acronym{nuSMEFT}{Right-handed Neutrino extended Standard Model effective field theory}[Right-handed Neutrino extended Standard Model effective field theories]
\acronym{HEFT}{Higgs effective field theory}[Higgs effective field theories]
\acronym{LEFT}{light effective field theory}[light effective field theories]
\acronym{SCET}{soft-collinear effective theory}[soft-collinear effective theories]
\acronym{HQET}{heavy quark effective theory}[heavy quark effective theories]
\acronym{NRQCD}{non-relativistic quantum chromodynamics}
\acronym{NRQED}{non-relativistic quantum electrodynamics}
\acronym[2HDM]{THDM}{two Higgs-doublet model}
\acronym{SUSY}{supersymmetry}
\acronym{SUGRA}{supergravity}
\acronym{MSSM}{minimal supersymemtric Standard Model}
\acronym{NMSSM}{next-to-minimal supersymemtric Standard Model}
\longacronym{CC}{charged current}
\longacronym{NC}{neutral current}
\longacronym{LE}{low energy}[low energies]
\longacronym{HE}{high energy}[high energies]
\acronym{BR}{branching ratio}
\acronym{GW}{gravitational wave}
\acronym{GWB}{gravitational wave background}
\acronym{LISA}{Laser Interferometer Space Antenna}
\acronym{ET}{Einstein Telescope}
\acronym{DECIGO}{Deci-hertz Interferometer Gravitational wave Observatory}
\acronym{BBO}{Big Bang Observer}
\acronym{CMB}{cosmic microwave background}
\acronym{SGWB}{stochastic gravitational wave background}[stochastic gravitational wave backgrounds]

\acronym{SNF}{Swiss National Science Foundation}
\acronym{UCLouvain}{Université catholique de Louvain}
\acronym{FSR}{Fonds Spéciaux de Recherche}

\acronym[exp]{ex}{experimental}
\acronym{lat}{lattice}

\newcommand{\makediff}[2]{\expandafter\NewDocumentCommand\csname#1\endcsname{ogd()}{\IfNoValueTF{##2}{\IfNoValueTF{##3}{#2\IfNoValueTF{##1}{}{^{##1}}}{\mathinner{#2\IfNoValueTF{##1}{}{^{##1}}\argopen(##3\argclose)}}}{\mathinner{#2\IfNoValueTF{##1}{}{^{##1}}##2} \IfNoValueTF{##3}{}{(##3)}}}}
\makediff{mD}{\mathcal D}

\DeclareMathOperator{\U}{U}
\DeclareMathOperator{\sign}{sign}

\newcommand{\ii}{\mathrm{i}}

\makeatletter

\def\custom#1{{\hbox{$\left#1\vbox to30\p@{}\right.\n@space$}}}
\makeatother

\def\int{\intop\nolimits}

\graphicspath{{./figures/}}
\bibliography{bibliography}

\title{Thermal resummation for Very Strong First-order Phase Transitions}

\author[a]{Philipp Klose\orcidlink{0000-0003-3702-4738}%
\footnote{\href{mailto:pklose@nikhef.nl}{\texttt{pklose@nikhef.nl}}}}

\affil[a]{Nikhef, Theory Group, Science Park 105, 1098 XG, Amsterdam, The Netherlands}

\begin{document}

\maketitle 

\begin{abstract}
Effective potentials are a key ingredient for predicting stochastic gravitational wave backgrounds from strong first-order phase transitions in the early universe.
Established techniques for a robust computation, including dimensional reduction, rely on a high-temperature expansion
that is expected to break down for very strong transitions capable of producing observable backgrounds at next-generation gravitational wave detectors such as LISA.
We argue that existing 2PI effective action techniques enable consistent resummation for such transitions,
and use them to compute the next-to-leading order effective potential of the Abelian Higgs model for a strong transition in a general covariant gauge.
We find that our result can be recovered from a Daisy resummed potential by modifying the power counting,
and show explicitly that it satisfies the leading-order Nielsen identity needed for gauge-independent predictions of the bubble nucleation rate
and is consistent with prior results for small Higgs condensates.
\end{abstract}

\tableofcontents
\clearpage

\section{Introduction}

Phase transitions in the early universe plasma are an important source of \SGWBs that can be observed at next-generation interferrometers such as LISA.
The effective potential is the primary particle physics input needed to predict their amplitudes and spectra.
Importantly, in-medium effects such as Debye screening can signficantly affect the potential even at leading order \cite{Dolan:1973qd,Kirzhnits:1976ts}.
To include these effects, one has to resum the effective potential using an apropriate thermal resummation scheme.

Likely the most popular one is the so-called Daisy resummation scheme \cite{Parwani:1991gq,Arnold:1992rz},
which is attractive because it is easy to implement and provides a straight-forward physical interpretation as a thermal mass correction.
Unfortunately, it can induce undesireable artifacts, including an unphysical renormalization schale or gauge parameter dependence,
that induce large theory uncertainties for predictions of the SGWB \cite{Wainwright:2011qy,Croon:2020cgk}.
Dimenional reduction is a more robust scheme that addresses both of these issues and is suitable
for computing effective potentials in a wide range of models \cite{Farakos:1994xh,Kajantie:1996mn,Losada:1998at,Andersen:1998br,Niemi:2018asa,Gorda:2018hvi,Niemi:2020hto}.
It relies on a high-temperature expansion that is expected to break down for electroweak-type phase transitions that are strong enough to produce observable \SGWBs \cite{Chala:2024xll,Bernardo:2025vkz}.
Such transitions are of particular phenomenological interest because they are a generic feature of many standard model extensions that modify the effective Higgs potential, 
and because they can produce a large \SGWB in an ideal frequency range for obserevation at space-based interferrometers such as LISA \cite{LISA:2017pwj,LISACosmologyWorkingGroup:2022jok}.
Crucially, dimensional reduction is also the starting point for 3d lattice computations of the effective potential \cite{Farakos:1994xh,Laine:1995np,Kajantie:1996mn,Kajantie:1996qd,Laine:1997dy,Laine:1998qk,Laine:2012jy}.
Hence, these computations rely on the same high-temperature expansion and also become unreliable for very strong transitions.
Direct 4d lattice simulations are not subject to the same limitation, and were previously used to study first-order phase transtions in relatively simple Higgs-gauge theories \cite{Fodor:1999at}.
However, such simulations are not feasable for theories that contain chiral fermions with non-Abelian chiral gauge interactions \cite{Luscher:2000hn,Karsch:2001cy,HotQCD:2014kol}.
Therefore, there is a need for a robust resummation scheme that does not produce unphysical artifacts while
remaining applicable for very strong transitions.
\\

One alternative to dimensional reduction is to resum the effective potential by solving gap equations to obtain the correct in-medium dispersion relation of each particle species.
This approach is used in the partial dressing scheme \cite{Boyd:1993tz,},
which is relatively simple and produces the correct leading-order effective potential
in simple models without fermions or gauge symmetries \cite{Curtin:2016urg,Curtin:2022ovx}.
The main drawback of this scheme is that its gap equations are not derived from first principles, which makes it difficult to generalize results beyond leading order.
It is also not clear whether it produces correct results when applied to more complicated theories with \eg spontaneously broken gauge symmetries.
Both issues can be solved by using a 2PI effective action approach \cite{Cornwall:1974vz,Berges:2015kfa} to derive consistent gap equations from first principles of quantum field theory.
This approach was first used to resum the effective potential of a $\phi^4$ theory in \cite{Amelino-Camelia:1992qfe},
and has also been applied to the finite temperature potential of the Abelian Higgs model \cite{Amelino-Camelia:1993rvt}. 
It was also used recently to compute the describe the dynamics of a confining transition in chiral perturbation theory \cite{Houtz:2025ogg}.

The main issue is that the 2PI approach is technically challenging and therefore difficult to implement in practice.
It has also never been demonstrated that it actually yields gauge-independent predictions for observables such as the phase transition strength or the bubble nucleation rate.
To our knowledge, the old computation in \cite{Amelino-Camelia:1993rvt} is the only existing attempt to use 2PI effective action techniques
for computing the finite-temperature effective potential in a spontaneously broken gauge theory.
This computation still employs a strict high-temperature expansion to solve the gap equations, which means that is not suitable for very strong phase transitions,
and is performed in Landau gauge instead of a general covariant gauge, which makes it impossible to test gauge-parameter independence.

The main goal of the present work is to explictly demonstrate that 2PI effective action techniques can be used
to compute resummed effective potentials governing strong first-order phase transitions in spontaneously broken gauge theories
that yield gauge-parameter independent predictions for physical observables.
To this end, we develop a new method for computing effective potentials in the real-time approach
that makes it possible to resum the potential by leveraging various existing results for the 2PI effective action.
We then use this method to compute the effecte potential for a strong transition of the Abelian Higgs model in a general covariant gauge,
and show that it satisfies the Nielsen identity that is needed for gauge-parameter independent predictions of the bubble nucleation rate \cite{Metaxas:1995ab,Hirvonen:2021zej,Lofgren:2021ogg}.
%This improves the old computation in \cite{Amelino-Camelia:1993rvt}
% and provides the first ab-initio computation of a resummed potential for very strong cosmological phase transitions in a Higgs gauge theory.
\\

The remainder of the article is structured as follows:
In \cref{sec:2pi potentials},
We will present a general method for computing 2PI resummed effective potentials in the real-time formalism.
In \cref{sec:higgs gauge theories}, we then apply this method to compute the next-to-leading order
effective potential for a strong first-order phase transition in the Abelian Higgs model.
Finally, we show that the potential satisfies the Nielsen identity needed for gauge-parameter independent predictions
and that it reproduces existing results from dimensional reduction for small Higgs-condesates.
\cref{sec:conclusion} concludes the article.
Finally, \cref{app:nielsen,app:photon spectral} provide additional details for deriving
the Nielsen identity and the structure of the resummed photon spectral function.

\section{Resummed effective potentials in the real-time formalism}
\label{sec:2pi potentials}

In this section, we present a general approach for computing 2PI resummed effective potentials in the real time formalim.
This approach shares features with the one pursued in the older works \cite{Amelino-Camelia:1992qfe,Amelino-Camelia:1993rvt},
which also use 2PI techniques to compute effective potentials in the imaginary time formalism.
We instead adopt a real-time approach because that makes it easy to employ various pre-existing results for the 2PI effective action,
which is most commonly used as a tool for real time computations.
To prepare the susequent discussion, we first review the usual definition of the effective potential in the imaginary time formalism
before moving on to the compution in the real-time approach.
We then discuss how to resum the potential using 2PI effective action techniques.
See also \cite{Laine:2016hma} for a review of the usual imaginary time computation,
\eg \cite{Cornwall:1974vz,Prokopec:2003pj,Berges:2015kfa} for an introduction to 2PI effective action techniques and the real-time approach,
and \cite{Garny:2008mig} for an overview of various technical subtleties associated with the 2PI effective action.
\\

Our ultimate goal is to determine the equilibrium properties,
including e.g. the latent heat and the statistical part of the bubble nucleation rate,
of first-order phase transitions in the early universe.
Phase transitions are non-analyticities of the free energy density
\begin{align}
\label{eq:free energy density}
f &\equiv \frac1{\beta V} \ln \tr( \rho_\text{eq} ) \ , &
\rho_\text{eq} &= e^{- \beta H} \ ,
\end{align}
where $\rho_\text{eq}$ is the equilibrium von-Neumann density matrix, $\beta = 1/T$ the corresponding inverse temperature, $V$ the spatial volume,
and $H$ the Hamiltonian of the relevant theory.
In the early universe, these non-analyticities are typically associated with a discontinuous condensate
\begin{align}\label{eq:condensate def}
\varphi_\text{eq} &\equiv \tr ( \rho_\text{eq} \, \Phi ) \ , &
\lim_{T \to T_c^+} \varphi_\text{eq} (T) \neq \lim_{T \to T_c^-} \varphi_\text{eq} (T) \ .
\end{align}
where $\Phi$ is some scalar field%
\footnote{If $\Phi$ carries a gauge charge, it is technically the gauge invariant condensate $\langle \Phi^\dagger \Phi \rangle$ that drives phase transitions.
In a perturbative computation, one typically introduces a gauge-fixing prescription and decomposes the complex field $\Phi$ into real components
that can then individually develop a condensate of the form given in \cref{eq:condensate def}.}
and $T_c$ the critical temperature associated with the transition.
In the imaginary time formalism, the traces in \cref{eq:free energy density,eq:condensate def} are written as path-integrals
over field configurations in $3+1$ dimensional space-time with a Euclidian metric and a compactified, circular time dimension with circumference $\beta$.
It can then be shown that the thermal equilibrium condensate minimizes the 1PI effective action
\begin{align}\label{eq:eucl eff act}
\Gamma^\text{E}_\text{1PI} [\varphi]
&= \int\displaylimits_{\substack{\text{1PI}\\\text{conn.}}} \hspace{-5pt} \mD \Phi_E \ \mD \mathcal X_{E \, i} \ e^{- S_E[\Phi_E + \varphi, \mathcal X_{E \, i} ]} \ ,
\end{align}
where $\Phi_E$ is the euclidian scalar field associated with $\varphi$,
$\mathcal X_{E \, i}$ collectively denotes any other euclidian fields, and $S_E$ is the corresponding euclidian classical action.
Since the euclidian time dimension is circular, bosons have to be periodic and fermions anti-periodic under time-translation with period $\beta$.
In particular, the euclidian scalar field obeys $\Phi_E(\tau, \bm x) = \Phi_E (\tau + \beta, \bm x)$.

By construction, the condensate $\varphi$ in \cref{eq:condensate def} is a time-independent constant.
The effective action is therefore proportional to a trivial volume factor $\beta V$, and the effective potential
\begin{align}
V_\text{eff} (\varphi ) &= \frac1{\beta V} \times \Gamma^\text{E}_\text{1PI} [\varphi]
\end{align}
is defined by removing this factor.
Since the potential is a simple function of $\varphi$, its local minima are solutions of the equation
\begin{align} \label{eq:minimal potential}
0 &= V_\text{eff}^\prime (\varphi) \ .
\end{align}
Importantly, the effective potential can exhibit multiple local mimima, each corresponding to a distinct phase of the primordial plasma.
The most energetically favourable one, i.e. the global minimum, determines the ``true'' equilibrium value of the condensate $\varphi_\text{eq}$ in the stable phase,
while the other minima determine the value of the condensate in any additional meta-stable phases.
The free-energy density defined in \cref{eq:free energy density} is equal to the value of the effective potential in the global minimum,
\begin{align}
f &= V_\text{eff} (\varphi_\text{eq}) \ .
\end{align}
As the temperature of the primordial plasma changes, the previously stable phase can become unstable.
Depending on the shape of the effective potential, this results in either a phase transition or an analytic cross-over to the new global minimum.
In many cases, the potential, and with it the core properties of the transition, is sensitive to in-medium effects that modify the propagation of long-range correlations.
It is then crucial to include these effects by using an appropriate thermal resummation scheme.

\subsection{Real time approach}

To set up the computation of the effective potential in a real-time formalism, we first have to make contact with its original definition in the imaginary time formalim.
The key difference between the two formalisms is that the real time approach does not assume from the start that the system is in thermal equilibrium.
This means that the condensate $\varphi(t)$ may be time-dependent.
However, if the system is thermodynamically stable, it has to equilibrate at late times,
\begin{figure}
\begin{align*}
\bar \Gamma_2
&\supset \underbrace{%
\includegraphics[height = .1\textwidth, valign = c, clip, trim = 0 0 0 0]{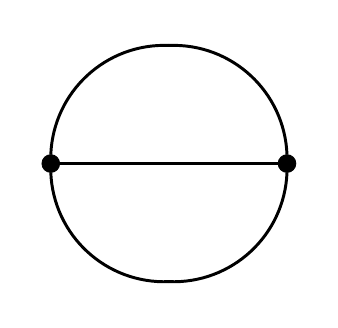}
+ \includegraphics[height = .1\textwidth, valign = c, clip, trim = 0 0 0 0]{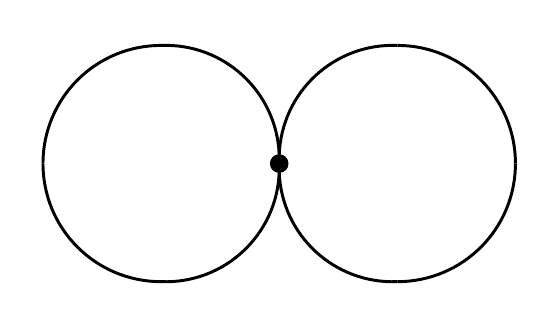}
+ \includegraphics[height = .1\textwidth, valign = c, clip, trim = 0 0 0 0]{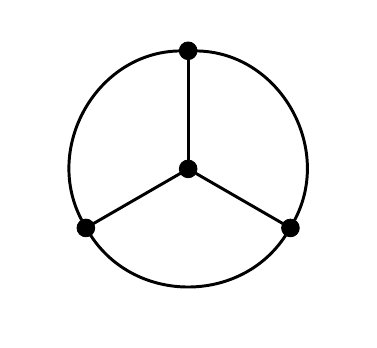}
+ \includegraphics[height = .1\textwidth, valign = c, clip, trim = 0 0 0 0]{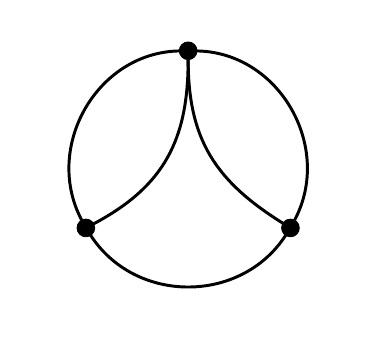}
+ \includegraphics[height = .1\textwidth, valign = c, clip, trim = 0 0 0 0]{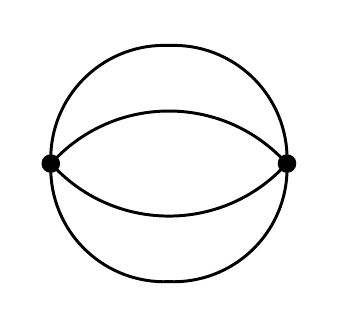}
}_\text{two-particle irreducible}
\\
&+ \underbrace{%
\includegraphics[height = .1\textwidth, valign = c, clip, trim = 0 0 0 0]{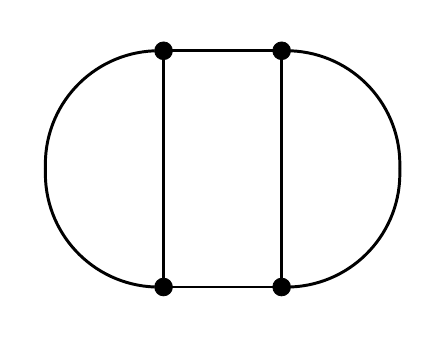}
+ \includegraphics[height = .1\textwidth, valign = c, clip, trim = 0 0 0 0]{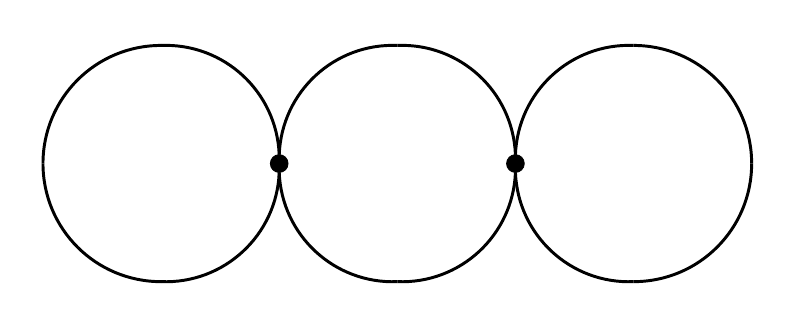}
+ \includegraphics[height = .1\textwidth, valign = c, clip, trim = 0 0 0 0]{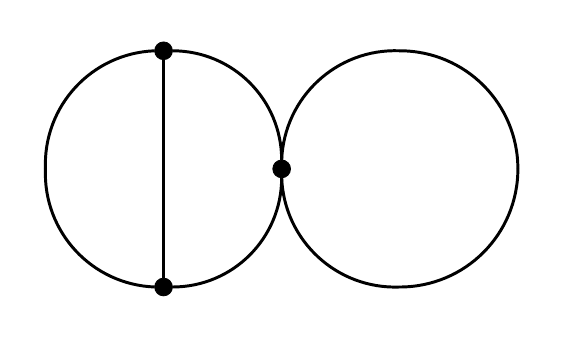}
}_\text{one-particle irreducible}
\end{align*}
\caption{\label{fig:1pi topologies}
Feynman diagram topologies with three or fewer loops that contribute to $\bar \Gamma_2$ and $\bar J_\varphi$ as defined in \cref{eq:1pi ctp action,eq:jbar def} in a renormalizeable theory.
The lines denote generic tree-level propagators. Diagrams in the first line are two-particle irreducible while diagrams in the second line
are only one-particle irreducible and therefore do not appear directly in either $\Gamma_2$ or $J_\varphi$ as defined in \cref{eq:2pi loop exp,eq:j def}.
}
\end{figure}
\begin{align}\label{eq:late time condensate}
\lim_{t\to \infty} \varphi(t) = \varphi_\text{eq} \ , 
\end{align}
so that we can extract information about thermal equilibrium by studying the late-time evolution of the condensate. 
Its equation of motion is determined by minimizing the real-time effective action
\begin{align}\label{eq:ctp 1pi action}
\ii \, \Gamma_\text{1PI}[\varphi^a] &= \int\displaylimits_{\substack{\text{1PI}\\\text{conn.}}} \hspace{-5pt} \mD \Phi^a \ \mD \mathcal X^a \ \langle t_0^+ | \rho | t_0^- \rangle \, e^{\ii \, S_\text{CTP}[\Phi^a + \varphi^a, \mathcal X_i^a]} \ ,
\end{align}
where the fields are now evaluated in real Minkowski space but along a closed time-path (CTP)
that starts at the initial time $t_0$ and runs to some final time in the far future, $t_f \to \infty$,
before returning back to the initial time $t_0$.
Accordingly, each field picks up an index $a=\pm$ that denotes whether the field is evaluated along the forward (``$+$'') or backward (``$-$'') branch of the time path.
$S_\text{CTP}$ is the closed time-path action, and $\rho$ the von Neumann density matrix encoding the initial state, which does not have to be in equilibirum.
The matrix element
\begin{align}\label{eq:init matrix}
&\langle t_0^+ | \rho | t_0^- \rangle \ , &
| t_0^a \rangle &\equiv | \Phi^a (t_0, \bm x) \, \mathcal X_i^a (t_0, \bm x) \rangle
\end{align}
depends only on the field configurations at the forward and backward ends of the closed time-path.
If the system is time-independent with a thermal equilibrium inital state, the real-time effective action
encodes the same information as its euclidian counter-part \eqref{eq:eucl eff act}, and the condensate $\varphi(t) = \varphi_\text{eq}$ is constant.
It is in general quite difficult to directly compute the closed time-path effective action from its definition \eqref{eq:ctp 1pi action} because the intial state matrix element \eqref{eq:init matrix}
generates surface terms that are only finite on the initial time-slice $t = t_0$ but nevertheless contribute to the effective action and therefore need to be included for a self-consistent result.
By considering the late-time evolution of the system, we sidestep this issue.
In this regime, contributions due to the surface terms are strongly suppressed because
the dynamics of the system wash out information about its initial state.

In practice, we extract the effective potential from the equation of motion for $\varphi(t)$,
which has to be consistent with \cref{eq:late time condensate} in order to correctly predict the eventual equilibration of the condensate.
In particular, the equation of motion has to reproduce \cref{eq:minimal potential} for constant values of $\varphi$.
We can thus read off the field derivative of the effective potential $V_\text{eff}^\prime(\varphi)$ from the equation
and then recover the potential via the fundamental theorem of calculus,
\begin{align}\label{eq:potential def}
\Delta V_\text{eff} &\equiv V_\text{eff} (\varphi) - V_\text{eff} (0) = \int\displaylimits_0^\varphi \text{d} \varphi^\prime V_\text{eff}^\prime (\varphi^\prime) \ .
\end{align}
At a qualitative level, this stratey is similar strategy to the one used the partial dressing approach,
where the effective potential is computed by integrating a resummed expression for $V^\prime_\text{eff} (\varphi)$ \cite{Boyd:1993tz,Curtin:2016urg,Curtin:2022ovx}.
The key difference between that approach and our own is that the present approach makes it possible to compute the potential derivative from first principles of quantum field theory,
while the partial dressing formalism only postulates the shape of the gap equation used to resum $V^\prime_\text{eff} (\varphi)$.

In the remainder of this section, we derive an explicit formula for the potential derivative $V^\prime_\text{eff} (\varphi)$.
To this end, we consider an action of the shape
\begin{align}
S_\text{CTP} [\Phi^a, \mathcal X_i^a] &= \sum_a a \int \text{d}^4 x \left[ \frac12 \partial_\mu \Phi^a \partial^\mu \Phi^a - V_0 (\Phi^a) \right] + S_\text{CTP}^{\mathcal X} \ ,
\end{align}
where the bath action $S_\text{CTP}^{\mathcal X}$ depends on $\Phi$ and some number of additional $\mathcal X_i^a$ fields.
By using a loop expansion, one immediately obtains the standard expression
\begin{align}
\label{eq:1pi ctp action}
\Gamma_\text{1PI} [\varphi^a] &=
\underbrace{ \vphantom{\sum_s} S_\text{CTP} [\varphi^a, 0]}_\text{tree-level} +
\underbrace{ \sum_s \ii \, c_s \tr \ln( G_{0 \, s}^{-1} ) }_\text{1 loop} +
\underbrace{ \vphantom{\sum_s} \bar \Gamma_2 [\varphi^a]}_\text{2 or more loops} +
\underbrace{ \vphantom{\sum_s} \quad \text{surface terms}}_\text{from initial state} \ ,
\end{align}
where the explicit index $s$ distinguishes between distinct particle species,
and the constant $c_s$ depends on the type of each species (for example, one has $c_s = \nicefrac12$ for a real scalar field and $c_s = -1$ for a Dirac fermion).
The non-trivial contribution $\bar\Gamma_2$ captures the impact of plasma interactions and is equal to the sum of all one-particle irreducible diagrams.
\Cref{fig:1pi topologies} shows the Feynman diagram topologies that contribute to $\bar \Gamma_2$ in a renormalizeable theory at up to three loops.

Finally, $G_{0 \, s}^{-1}$ is the closed-time path equivalent of the zero-temperature inverse propagator $G_{0 \, s}^{-1}$.
It can carry gauge, flavour, Dirac, Lorentz, and closed time path indices and usually depends on the value of the condensate $\varphi^a$.
Accordingly, the functional trace $\tr ( \cdot )$ also runs over gauge, flavour, Dirac, Lorentz, and closed time path indices.
By using the action in \cref{eq:1pi ctp action}, one for example obtains the inverse $\Phi$ propagator
\begin{align}
[ \ii \, G_{0 \, \Phi}^{-1} ]^{ab} (x,y) &= - a \delta_{ab} \left( \partial_x^2 + V_0^{\prime\prime} (\varphi^a(x)) \right) \delta^4(x-y) \ .
\end{align}
For $t > t_0$, where the surface terms do not contribute directly, one then obtains the condensate equation of motion
\begin{align}\label{eq:1pi condensate eom}
0 &= - a \frac{\delta \Gamma_\text{1PI}}{\delta \varphi^a} 
= \ddot \varphi^a + V_0^\prime (\varphi^a) - \sum_s c_s \tr (G_{0 \, s} \otimes a \frac{\delta \, \ii \, G_{0 \, s}^{-1}}{\delta \varphi^a(x)}  ) - a \frac{\delta \Gamma_2 [\varphi]}{\delta \varphi^a(x)} \ ,
\end{align}
where
\begin{align}
(A \otimes B)^{ab}_{ij} (x,y) &= \sum_c c \int \text{d}^4 z \, A^{ac}_{ik} (x,z) B^{cb}_{kj} (z,y)
\end{align}
is a convolution with simultaneous matrix multiplication.
Here and in the following, the letters $i$, $j$, and $k$ symbollically denote any indices carried by the (inverse) propagator $G_{0 \, s}^{-1}$ and $G_{0 \, s}$.
To evaluate the functional trace in \cref{eq:1pi condensate eom}, one has to insert appropriate expressions for the propagators.
The variation of the inverse propagator is generically of the shape
\begin{align}\label{eq:inv prop var}
a \left[ \frac{\delta \, \ii \, G_{0 \, s}^{-1}(y,z)}{\delta \varphi^a(x)} \right]^{bc}_{ij} &= - \partial_\varphi A_{s \, ij} \, a b \, \delta_{ab} \delta_{bc} \delta^4(x-y) \delta^4 (y-z) \ ,
\end{align}
where the function $A_{s \, ij} (\varphi^a)$ is proportional to the mass-matrix of species $s$.
The functional trace then becomes
\begin{align}\label{eq:func trace}
- \tr (G_{0 \, s} \otimes a \frac{\delta \, \ii \, G_{0 \, s}^{-1}}{\delta \varphi^a}  ) &= 
\partial_\varphi A_{s \, ij} \ [G_{0 \, s}]^{aa}_{ij} (x,x) 
= \partial_\varphi A_{s \, ij} \int \frac{\text{d}^4 k}{(2\pi)^4} \, [G_{0 \, s}]^{aa}_{ij} (k) \ ,
\end{align}
where
\begin{align}
[G_{0 \, s}]^{ab}_{ij} (x, k) &= \int \text{d}^4 x \, e^{\ii k x} [G_{0 \, s}]_{ij}^{ab} \left( x + \frac12 r, x - \frac12 r \right)
\end{align}
is the Wigner-transformed tree-level propagator.
In thermal equilibrium, this propagator is independent of $x$ and reduces to the standard real-time propagator in momentum space.
The advanced and retarded combinations
\begin{align}\label{eq:ad ret propagators}
\ii G_{0 \, s}^{a} &= G_{0 \, s}^{++} - G_{0 \, s}^{-+} = G_{0 \, s}^{+-} - G_{0 \, s}^{--} \ , &
\ii G_{0 \, s}^{r} &= G_{0 \, s}^{++} - G_{0 \, s}^{+-} = G_{0 \, s}^{-+} - G_{0 \, s}^{--} 
\end{align}
then do not receive finite-temperature corrections and are the same as the advanced and retarted propagators at zero temperature.
The tree-level Wightman functions $G^{+-}$ and $G^{-+}$ obey the KMS relations
\begin{align}\label{eq:kms rel}
G_{0 \, s}^{-+} &= 2 h_s f_s \, \rho_{0 \, s} \ , &
G_{0 \, s}^{+-} &= 2 (1 + h_s f_s) \, \rho_{0 \, s} \ , &
\rho_{0 \, s} &= \frac1{2 \ii} \left(G_{0 \, s}^{a} - G_{0 \, s}^{r} \right) \ ,
\end{align}
where $\rho_{0 \, s}$ is the tree-level spectral function of species $s$ and
\begin{align}
f_s &= \frac1{e^{\beta k_0} - h_s} \ , &
h_s &= \pm 1
\end{align}
is either a Bose-Einstein or Fermi-Dirac distribution, depending of the spin of the particle species.
To separate the temperature-independent vacuum contributions to the effective potential from its finite temperature corrections,
we use that the distribution function
\begin{align}\label{eq:low t limit}
f_s (k_0) \xrightarrow{\beta \to \infty} - h_s \Theta(-k_0) \ ,
\end{align}
is proportional to a Heavyside function at zero temperature.  

Taken together, \cref{eq:func trace} and \crefrange{eq:ad ret propagators}{eq:low t limit} provide the ingredients we need
to extract the derivative of the effective potential from the condensate equation of motion \eqref{eq:1pi condensate eom}.
For a constant condensate, and also using that any physical solution to \cref{eq:1pi condensate eom} has to obey the constraint $\varphi^+ = \varphi^-$,
the equation reduces to the general formula
\begin{align}\label{eq:1pi pot}
V_\text{eff}^\prime (\varphi) &= V^\prime_0 (\varphi) + V^\prime_\text{cw} (\varphi) + \bar J_\varphi
+ \left[ 2 \sum_s h_s c_s \, \partial_\varphi A_{s \, ij} \int_k \, \left( f_s (k_0) + h_s \Theta(-k_0) \right) [\rho_{0 \, s}]_{ij} (k) \right]_\text{$\varphi =$ const.} \ .
\end{align}
The temperature-independent contribution
\begin{align}\label{eq:treelvl cw pot}
V^\prime_\text{cw} (\varphi)
&= \left[ \sum_s c_s \, \partial_\varphi A_{s \, ij} \int_k \, \sign(k_0) [\rho_{0 \, s}]_{ij} (k) \right]_\text{$\varphi =$ const.} 
\end{align}
is equal to the usual one-loop Coleman-Weinberg potential \cite{Coleman:1973jx},
while the current
\begin{align}\label{eq:jbar def}
\bar J (\varphi) &= \frac12 \left. \left( \frac{\delta \bar \Gamma_2}{\delta \varphi^+(x)} - \frac{\delta \bar \Gamma_2}{\delta \varphi^-(x)} \right) \right|_{\varphi^\pm = \text{\ const.}} 
\end{align}
captures higher-order loop corrections to the effective potential.

\subsection{2PI resummation}
\label{sec:2pi resum}

\begin{figure}
\begin{equation*}
\Gamma_2 \supset
\underbrace{%
\includegraphics[height = .1\textwidth, valign = c, clip, trim = 0 0 0 0]{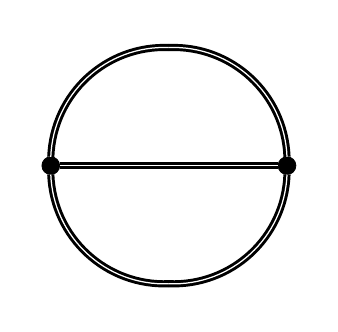}
+ \includegraphics[height = .1\textwidth, valign = c, clip, trim = 0 0 0 0]{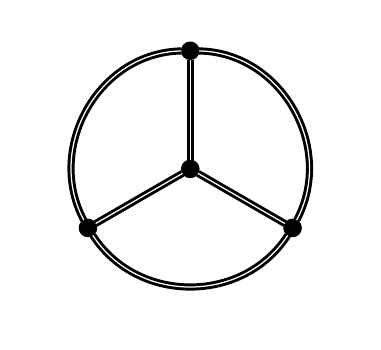}
+ \includegraphics[height = .1\textwidth, valign = c, clip, trim = 0 0 0 0]{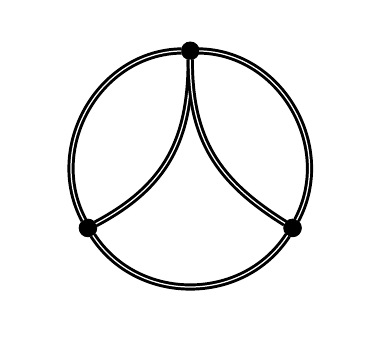}
}_\text{contribute to $J_\varphi$}
+ \includegraphics[height = .1\textwidth, valign = c, clip, trim = 0 0 0 0]{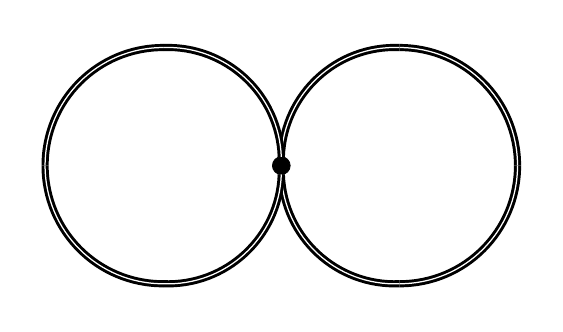}
+ \includegraphics[height = .1\textwidth, valign = c, clip, trim = 0 0 0 0]{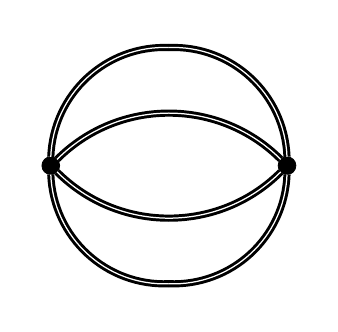}
\end{equation*}
\caption{\label{fig:2pi topologies}
Feynman diagram topologies with three or fewer loops that contribute to $\Gamma_2$ and $J_\varphi$ as defined in \cref{eq:2pi loop exp,eq:j def} in a renormalizeable theory.
The double lines denote exact propagators as defined in \cref{eq:ctp propagator}.
Only two-particle irreducible diagrams with at least one three-point vertex can contribute to $V_\text{eff}^\prime(\varphi)$,
which significantly reduces the number of topologies that have to be included at each order in perturbation theory.
}
\end{figure}

So far, we have not yet included any resummation.
We now use 2PI effective action techniques to consistently resum the potential derivative $V_\text{eff}^\prime$.
The main idea is to treat the exact two-point functions
\begin{align}\label{eq:ctp propagator}
G_\Phi^{ab} (x,y) &\equiv \langle T_C \{ \Phi^a(x) \Phi^b(y) \} \rangle - \varphi^a(x) \varphi^b(y) \ , &
[G_s]_{ij}^{ab} (x,y) &= \langle \langle T_C \{ \mathcal X_i^a(x) \mathcal X_j^b (y) \} \rangle \ ,
\end{align}
where $T_C$ is a closed time-path ordering operator, as independent degrees of freedom with individual equations of motion.
By solving these equations, we obtain resummed propagators that determine the correct in-medium dispersion relation of each particle,
and that we can use to compute the resummed potential.

To derive the propagator equations of motion, one has to construct an effective action that captures the coupled dynamics of the condensate $\varphi$ and the two-point functions \eqref{eq:ctp propagator}.
Indeed, it can be shown that the physical values of $\varphi$ and $G$ extremize the 2PI effective action
\begin{align}
\ii \, \Gamma_\text{2PI} [\varphi, G_s] &= 
\int\displaylimits_{\substack{\text{2PI}\\\text{conn.}}} \hspace{-5pt} \mD \Phi^a \ \mD \mathcal X^a \ \left[ \langle t_0^- | \rho_\text{eq} | t_0^+ \rangle \, e^{\ii \, S_C[\Phi^a + \varphi^a, \mathcal X_i^a]} \right]_{G_{0 \, s} \to \, G_s} \ ,
\end{align}
where the subscript denotes that only two-particle irreducible diagrams contribute to the path integral.
The condensate and correlation functions therefore obey the coupled equations of motion
\begin{align}\label{eq:2pi eoms}
0 &= \frac{\delta \Gamma_\text{2PI}[\varphi, G]}{\delta \varphi^a(x)} \ , &
0 &= \frac{\delta \Gamma_\text{2PI}[\varphi, G]}{\delta [G_s]_{ji}^{ba} (y,x)} \ .
\end{align}
The first equation is obtained by varying the action with respect to $\varphi$ and corresponds to \cref{eq:1pi condensate eom} in the 1PI approach.
We will use it to derive our expression for $V_\text{eff}^\prime$ in terms of the resummed propagators $G_s$.
They are solutions to the secondary equation on the right-hand side, which is obtained varying the action with respect to $G_s$.
Importantly, the resummed propagators still obey the KMS relations in \cref{eq:kms rel}, which means that only the equation for the spectral function is non-trivial.
Furthermore, results for the resummed spectral function are well known for many relevant models.
This means that it is rarely necessary to actually solve the propagator equations of motion.

To obtain our formula for $V_\text{eff}^\prime$, we follow the same steps as in case of the 1PI effective action.
A perturbative loop expansion of the 2PI effective action yields the standard expression
\begin{align}\label{eq:2pi loop exp}
\Gamma_\text{2PI} [\varphi, G] &= S_\text{CTP}[\varphi] + \sum_s \ii \, c_s \left( \tr \ln (G_s^{-1}) + \tr (G_{0 \, s}^{-1} G_s) \right) + \Gamma_2 [\varphi, G] + \text{surface terms} \ ,
\end{align}
where $\Gamma_2$ is now the non-trivial interacting part of the 2PI effective action, which is given as a sum of two-particle irreducible diagrams.
\Cref{fig:2pi topologies} shows the Feynman diagram topologies that contribute to $\Gamma_2$ at up to three loops.
Note that each diagram is constructed by using the \emph{exact} propagators defined in \cref{eq:ctp propagator} instead of the usual tree-level propagators.
After following the same steps as before, one obtains the derivative of the effective potential
\begin{align}\label{eq:2pi pot}
V_\text{eff}^\prime (\varphi) &= V^\prime_0 (\varphi) + V^\prime_\text{cw} (\varphi)
+ \left[ 2 \sum_s h_s c_s \, \partial_\varphi A_{s \, ij} \int_k \, \left( f_s (k_0) + h_s \Theta(-k_0) \right) [\rho_{s}]_{ij} (k) \right]_\text{$\varphi =$ const.} 
+ J_\varphi \ ,
\end{align}
where
\begin{align} \label{eq:resummed cw}
V^\prime_\text{cw} (\varphi)
&= \left[ \sum_s c_s \, \partial_\varphi A_{s \, ij} \int_k \, \sign(k_0) [\rho_s]_{ij} (k) \right]_\text{$\varphi =$ const.} 
\end{align}
is the resummed equivalent to the unresummed Coleman-Weinberg potential in \cref{eq:treelvl cw pot}.
The main difference between the two is that \cref{eq:resummed cw} depends on the resummed spectral functions
\begin{align}
\rho_s &= \frac1{2\ii} \left( G_s^a - G_s^r \right) = \frac1{2} \left( G_s^{+-} - G_s^{-+} \right) \ .
\end{align}
This implies that the resummed Coleman-Weinberg potential picks up thermal contributions that are generated by finite-temperature corrections to the dispersion relation of each particle.
The current
\begin{align}\label{eq:j def}
J (\varphi) &= \frac12 \left. \left( \frac{\delta \Gamma_2}{\delta \varphi^+(x)} - \frac{\delta \Gamma_2}{\delta \varphi^-(x)} \right) \right|_{\varphi^\pm = \text{\ const.}} 
\end{align}
again captures higher-order loop corrections. The underbrace in \cref{fig:2pi topologies} shows the topologies that contribute to $J (\varphi)$ in a renormalizeable theory.
The remaining diagrams still contribute indirectly because they enter into the computation of the resummed spectral functions.
Nevertheless, the number of diagrams contributing to $J (\varphi)$ is much smaller than in the number of diagrams contributing to the 1PI current $\bar J(\varphi)$,
which simplifies the inclusion of higher order corrections and is expected to improve the convergence of the loop expansion.
\\

The formula in \cref{eq:2pi pot} is the main result of this section.
By combing it with the propagator equations on right-hand side of \cref{eq:2pi eoms}
and the fundamental relation between $V^\prime(\varphi)$ and $V(\varphi)$ in \cref{eq:potential def},
we have now assembled all the ingredients needed to compute the effective potential.
We found that it can be computed by following a straightforward step-by-step recipe:
\begin{enumerate}
\item
Construct the 2PI effective action by including all relevant Feynman diagrams.
Only two-particle irreducible diagrams contribute to the action, and each propagator corresponds to an \emph{exact} two-point correlation function rather than its tree-level approximation.
\item
Use the propagator equations of motion on the right-hand side of \cref{eq:2pi eoms} to compute resummed spectral functions for each particle species.
In practice, the existing literature already provides many results for thermally resummed spectral functions.
\item
Insert the resummed spectral functions into the loop expanded formula \eqref{eq:2pi pot} to compute $V^\prime_\text{eff}$.
At this step, one has to renormalize the Coleman-Weinberg contribution to the effective potential and any other contributions that do not vanish at zero temperature.
\item
Integrate the result for $V^\prime_\text{eff}$ to compute $\Delta V_\text{eff}$.
\end{enumerate}
No matter which method is used to compute the effective potential, it is crucial
to impose a consistent power counting in each step of the computation \cite{Lofgren:2023sep}.
For the method presented here, this implies that one has to employ the same counting
when evaluating the spectral function and when using \cref{eq:2pi pot} to compute the derivative of the effective potential.
Crucially, a naive loop expansion may not provide a consistent power counting if it mixes different orders in a strict small coupling expansion.
An inconsistent computation can induce unphysical gauge dependences for key observables such as the bubble nucleation rate \cite{Wainwright:2011qy,Croon:2020cgk}.
This makes tests of gauge-independence an important tool to help ensure the consistency of a chosen power counting.

\section{Abelian Higgs Model}
\label{sec:higgs gauge theories}

We now use the resummation procedure presented in the previous section to
compute the effective potential for a strong first-order phase transition in the Abelian Higgs model.

While this model is relatively simple, it exhibits many of the key features, including a radiatively generated barrier and the need for a careful treatment of gauge bosons,
that complicate the computation of the effective potential in more general Higgs-gauge theories.
As a gauge theory, it can also give rise to unphysical gauge-parameter dependencies if the power counting used to compute the potential is inconsistent.
It is therefore an ideal proving-ground to demonstrate the efficacy of the resummation procedure.
The Lagrangian of the Abelian Higgs model is given as
\begin{align}\label{eq:abelian higgs model}
\mathcal L &= (D^\mu \phi)^\dagger D_\mu \phi + \mu^2 (\phi^\dagger \phi) - \lambda (\phi^\dagger \phi)^2 - \frac14 F_{\mu\nu} F^{\mu\nu} + \mathcal L_\text{gf} + \mathcal L_\text{ghost} \ ,
\end{align}
where $\phi$ is the complex Higgs field, $\mu^2 > 0$ a mass parameter,
$D_\mu = \partial_\mu + \ii g A_\mu$ its covariant derivative,
$A_\mu$ a $\U(1)$ gauge field, $g$ the corresponding coupling constant, and
\begin{align}
F_{\mu\nu} &= \partial_\mu A_\nu - \partial_\nu A_\mu 
\end{align}
the field strength tensor.
The gauge-fixing and ghost Lagrangians are
\begin{align}
\mathcal L_\text{gf} &= \frac1{2 \xi} F^2 \ , &
\mathcal L_\text{ghost} &= \overline c \left( \left. \frac{\delta F}{\delta \alpha} \right|_{\alpha = 0} \right) c \ ,
\end{align}
where $F$ is a gauge-fixing functional, $\overline c$ and $c$ are ghost fields,
and the scalar function $\alpha(x)$ parametrizes the gauge transformation
\begin{align}
\phi &\to e^{\ii g \alpha} \phi \ , &
A_\mu &\to A_\mu - \partial_\mu \alpha \ . 
\end{align}
To compute the effective potential, we decompose the Higgs field as
\begin{align}
\phi &= \frac1{\sqrt2} \left( \varphi + h + \ii \chi \right) \ ,
\end{align}
where $\varphi$ is the Higgs condensate, $h$ the physical Higgs boson, and $\chi$ the unphysical Goldstone boson.
We work in a general $R_\xi$ gauge, choosing the gauge-fixing functional
\begin{align}\label{eq:gauge fix}
F &= - \partial_\mu A^\mu - \xi m_A \, \chi \ , &
\left. \frac{\delta F}{\delta \alpha} \right|_{\alpha = 0} &= - \left[ \partial^2 + m_c^2 \left( 1 + \frac{h}{\varphi} \right) \right] \ ,
\end{align}
where
\begin{align}
m_A &= g \, \varphi &
\text{and} &&
m_c^2 &= \xi m_A^2
\end{align}
are the gauge boson and ghost field mass parameters.
Notice that this prescription differs from the one used in \cite{Amelino-Camelia:1993rvt}.
That work also used an effective action approach to compute the effective potential in the Abelian Higgs model,
but chose the gauge-fixing functional
\begin{align}\label{eq:alt gauge}
F_\text{alt.} &= - \partial_\mu A^\mu \ , &
\left. \frac{\delta F_\text{alt.}}{\delta \alpha} \right|_{\alpha = 0} &= - \partial^2 \ .
\end{align}
This prescription is simpler than the one used here because it causes the ghost fields to completely decouple from the remainder of the theory
and because the tree-level mass parameter of the Goldstone boson field $\chi$ is then gauge-parameter independent.
The cost is that the gauge-fixing in \cref{eq:alt gauge} leads to a mixing term
\begin{align}
\mathcal L \supset - m_A \, \chi \, \partial_\mu A^\mu 
\end{align}
that has be treated perturbatively.
This is not an issue in the regime $m_A \sim g T$, where dimensional reduction is also applicable, but problematic for large masses $m_A \sim \pi T$.
Since this regime is our primary focus, we have to use the more complicated prescription defined in \cref{eq:gauge fix}.
One then obtains the tree-level propagators
\begin{subequations}\label{eq:tree invprop}
\begin{align}
[ \ii G_{0 \, h} ]^{-1} (x,y) &= - \left( \partial^2_x + m_h^2 \right) \delta^4 (x-y)
\ , \\
[ \ii G_{0 \, c} ]^{-1} (x,y) &= - \left( \partial^2_x + m_c^2 \right) \delta^4 (x-y)
\ , \\
[ \ii G_{0 \, \chi} ]^{-1} (x,y) &= - \left( \partial^2_x + m_\chi^2 \right) \delta^4 (x-y)
\ , \\
[ \ii G_{0 \, A}]^{-1} _{\mu\nu} (x,y) &= - \left( \left( 1 - \frac1{\xi} \right) \partial_{x\,\mu} \partial_{x\,\nu} - g_{\mu\nu} (\partial_x^2 + m_A^2) \right) \delta^4 (x-y)
\ ,
\end{align}\end{subequations}
where
\begin{align}
m_h^2 &= - \mu^2 + 3 \lambda \varphi^2 \ , &
m_\chi^2 &= - \mu^2 + \lambda \varphi^2 + m_c^2 
\end{align}
are the Higgs and Goldstone boson mass parameters.
In momentum space, the resulting tree-level spectral functions are given as
\begin{subequations}\label{eq:tree spec}
\begin{align}
\rho_{0 \, s} &= \sign(k_0) \pi \delta(k^2 - m_s^2) \ , &
s &= h, c, \chi
\end{align}
and
\begin{align}
[ \rho_{0 \, A} ]^{\mu\nu} &= - \left( \mathds P_T^{\mu\nu} + \mathds P_L^{\mu\nu} \right) \sign(k_0) \pi \delta(k^2 - m_A^2)  - \mathds P_D^{\mu\nu} \, \xi \, \rho_{0 \, c}
\ ,
\end{align}\end{subequations}
where we have defined the usual transverse and longitudinal projectors
\begin{align}\label{eq:proj def}
\mathds P^T_{\mu\nu} &= \delta_\mu^{\ i} \delta_\nu^{\ j} \left( g_{ij} + g_{00} \frac{k_i k_j}{\bm k^2} \right) \ , &
\mathds P^L_{\mu\nu} &= \frac{n_\mu n_\nu}{n^2} \ , &
\mathds P^D_{\mu\nu} &= \frac{k_\mu k_\nu}{k^2} \ , &
n_\mu &= g_{0\mu} - \frac{k_0 k_\mu}{k^2} \ .
\end{align}
They are pairwise orthogonal and obey the sum rules
\begin{align}\label{eq:proj sum rules}
\mathds P_T^{\mu\nu} + \mathds P_L^{\mu\nu} &= g^{\mu\nu} - \frac{k^\mu k^\nu}{k^2} \ , &
g_{\mu\nu} \mathds P_T^{\mu\nu} &= 2 \ , &
g_{\mu\nu} \mathds P_L^{\mu\nu} &= 1 \ , &
g_{\mu\nu} \mathds P_D^{\mu\nu} &= 1 \ .
\end{align}
At zero temperature, the Abelian Higgs model defined by \cref{eq:abelian higgs model} develops a non-vanishing condensate that spontaneously breaks its gauge symmetry.
Thermal effects stabilize then the symmetric phase by generating a potential barrier that turns the origin $\varphi = 0$ into a secondary local minimum.

Below the critical temperature $T_c$, the broken phase minimum at $\varphi = \varphi_\text{min}$ remains energetically favourable and the gauge symmetry is still broken.
Above this tempature, the minimum at the origin is energetically favourable, restoring the $\U(1)$ gauge symmetry.
Our goal is to study the properties of this transition in cases that are beyond the regime of applicability for dimensional reduction.
To do so, one has to compute the effective potential in the vicinity of the critical temperature $T_c$.

\subsection{Power counting}
\label{subsec:pwr count}

The first step in computing the effective potential is to define a consistent power counting for the perturbative loop expansion.
We primarily focus on cases where the strict high-temperature expansion underlying dimensional reduction breaks down.
This is the case if the Higgs-contribution to the gauge boson mass $m_A = g \, \varphi$ is similar to the hard scale $\pi T$ \cite{Bernardo:2025vkz}.
Since the broken phase minimum $\varphi_\text{min}$ is the largest relevant value of the condensate,
dimensional reduction is inconsistent for
\begin{align}\label{eq:phimin scaling}
\varphi_\text{min} &\gtrsim \frac{\pi T}{g} \ .
\end{align}
Besides the temperature and the gauge coupling $g$, the most imporant parameter that determines $\varphi_\text{min}$ is the ratio $x = \nicefrac{\lambda}{g^2}$ \cite{Ekstedt:2024etx}.
Assuming that $\varphi_\text{min}$ is small enough to justify a high-temperature expansion,
we can use dimensional reduction to construct an effective theory of soft modes with momenta $k \lesssim g T$ \cite{Kajantie:1995dw,Hirvonen:2021zej}.
A numerical lattice simulation then reveals the existence of a critical endpoint at $x_c \approx 0.28$, which implies that the theory is non-perturbative for $x > x_c$ \cite{Mo:2001fi}.
In the perturbative regime $x < x_c$,  one obtains the approximate minimum \cite{Ekstedt:2024etx}.
\begin{align}\label{eq:phimin dimred}
g \, \varphi_\text{min} &\approx \frac{g^2 T}{2\pi x} \ .
\end{align}
By comparing \cref{eq:phimin scaling,eq:phimin dimred}, we find that dimensional reduction is unreliable for
\begin{align}
x &\lesssim \frac{g^2}{\pi^2} &
&\Rightarrow &
\lambda &\lesssim g^4 \ .
\end{align}
This counting differs from the $g^4 \ll \lambda \ll g^2$ power counting that usually governs thermal phase transitions in Higg-gauge theories \cite{Arnold:1992rz,Farakos:1994kx,Hirvonen:2021zej}.
It is instead similar to the counting that governs radiative symmetry breaking at zero temperature,
where $\lambda$ also scales as $g^4$ \cite{Coleman:1973jx,Metaxas:1995ab,Hirvonen:2021zej}.

To compute the effective potential, we also have to determine the relative size of the mass parameter $\mu^2$. 
Since the two potential minima are approximately degenerate close to the critical temperature $T_c$, one finds that
\begin{align}\label{eq:strong counting}
\mu^2 &\sim g^4 \varphi_\text{min}^2 \sim g^2 \, T^2 &
&\text{for} &
T &\sim T_c \ .
\end{align}
This means that we can still treat $\mu^2$ as a small parameter, even in cases where $\varphi$ is compareble to the temperature.
Likewise, the mass difference between the Goldstone boson $\chi$ and the ghost fields 
\begin{align}
m_\chi^2 - m_c^2 &= - \mu^2 + \lambda \varphi^2 
\end{align}
is small even if the individual masses $m_\chi^2$ and $m_c^2$ are large.
In other words, we can perform a partial high-temperature expansion in which $m_h^2$ and $m_\chi^2 - m_c^2$ are treated as small expansion parameters
while $m_A^2$, $m_c^2$ and $m_\chi^2$ are large.
\\

By using the scaling of $\lambda$ and $\mu^2$,
we can now determine the scaling of the tree-level contributions to the effective potential in the vicinity of the broken phase minimum $\varphi_\text{min}$.
Explicitly, one finds
\begin{align}
\Delta V_0 (\varphi_\text{min}) &= - \frac12 \mu^2 \varphi_\text{min}^2 + \frac14 \lambda \varphi_\text{min}^4 \sim g^4 \varphi_\text{min}^4 \ , &
V^\prime_0 (\varphi_\text{min}) &\sim g^4 \varphi_\text{min}^3 \ .
\end{align}
In this regime, the large value of the gauge boson mass parameter $g \, \varphi_\text{min} \gtrsim \pi T$ implies
that thermal resummation is not necessary to compute the effective potential.
Since the transition strength
\begin{align}
\alpha(T) &\propto
T \frac{\text{d}}{\text{d} T} \frac{\Delta V(\varphi_\text{min})}{T^3}
\end{align}
only depends on the values of the potential at its minima,
there is no need for thermal resummation to compute the transition strength.

The situation is more complicated for the bubble nucleation rate,
which depends on the entire range of values of $V(\varphi)$ between the two minima and in particular on the the size of the barrier that separates them.%
\footnote{%
Besides the effective potential, the bubble nucleation also depends on wave-function type corrections to the kinetic term of the nucleation action and a fluctuation determinant, \cf \eg \cite{Hirvonen:2021zej,Kierkla:2025qyz} for details.
We here focus on the potential, and leave the computation of these quantities to future work.}
To compute the rate, one thus has to find the value of the potential for soft condensates
\begin{align}
\varphi_\text{soft} < \pi T \sim g \, \varphi_\text{min} \ .
\end{align}
In this regime, the leading tree-level contributions to the effective potential scale as
\begin{align}
\Delta V_0 (\varphi_\text{soft}) &= - \frac12 \mu^2 \varphi_\text{soft}^2 \sim g^2 \varphi_\text{soft}^4 \ , &
V^\prime_0 (\varphi_\text{soft}) &\sim g^2 \varphi_\text{soft}^3 \ .
\end{align}
To obtain the complete leading order potential, one therefore has
to include all contributions that scale like $g^4 \varphi_\text{min}^4 \sim T^4$
in the hard mass regime 
and like $g^2 \varphi_\text{soft}^4 \sim g^2 T^4$
in the soft mass regime,
\begin{align}\label{eq:lo pwr count}
\Delta V_\text{LO} &\sim
\begin{cases}
 T^4 & \varphi \sim \varphi_\text{min} \\
g^2 T^4 & \varphi \sim \varphi_\text{soft}
\end{cases} \ .
\end{align}
Higher order corrections have to be suppressed by additional powers of $g$ in both regimes.
At zero temperature and in the hard mass regime $g \varphi \sim T$, only terms with even powers contribute to the effective action,
so that the first beyond leading-order contribution scales as $g^2 V_\text{LO}$.
In the soft mass regime $\varphi \sim T$, finite temperature effects also generate terms that scale as odd powers of $g$.
Next-to-leading (NLO) contributions therefore scale as $g V_\text{LO}$ in the soft mass regime and as $g^2 V_\text{LO}$ in the hard mass regime,
while terms that are suppressed by a factor of $g^2$ in both regimes are formally of N\textsuperscript{2}LO,
\begin{align}\label{eq:nlo nnlo pwr}
\Delta V_\text{NLO} &\sim
\Delta V_\text{LO}
\begin{cases}
g^2 & \varphi \sim \varphi_\text{min} \\
g & \varphi \sim \varphi_\text{soft}
\end{cases} \ , &
\Delta V_\text{N\textsuperscript{2}LO} &\sim
\Delta V_\text{LO}
\begin{cases}
g^2 & \varphi \sim \varphi_\text{min} \\
g^2 &\varphi \sim \varphi_\text{soft}
\end{cases} \ .
\end{align}

\subsection{Effective potential}
\label{subsec:eff pot}

Our goal is to derive a consistent approximation of the effective potential
that is valid for any value of the condensate $\varphi$ and includes all LO and NLO contributions.
In particular, it has to generate gauge-parameter independent predictions for the phase transition strength and the bubble nucleation rate.
To achieve this, we use the resummation scheme presented in \cref{sec:2pi potentials}.

\begin{figure}
\centering
\begin{equation*}
V^\prime (\varphi) \supset
\underbrace{%
\includegraphics[height = .1\textwidth, valign = c, clip, trim = 0 0 0 0]{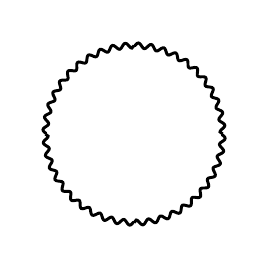}
+ \includegraphics[height = .1\textwidth, valign = c, clip, trim = 0 0 0 0]{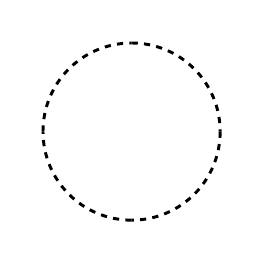}
+ \includegraphics[height = .1\textwidth, valign = c, clip, trim = 0 0 0 0]{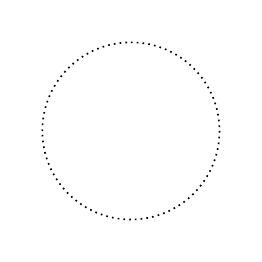}
}_\text{LO + NLO corrections}
+ \underbrace{%
\includegraphics[height = .1\textwidth, valign = c, clip, trim = 0 0 0 0]{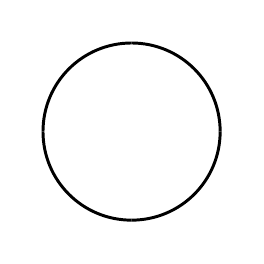}
+
\frac{\delta}{\delta \varphi} \left( \includegraphics[height = .1\textwidth, valign = c, clip, trim = 0 0 0 0]{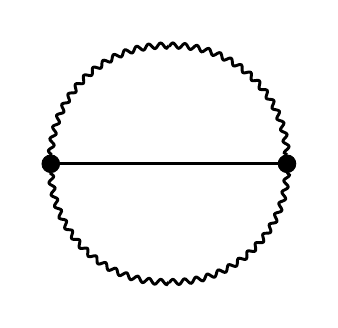}
+ \includegraphics[height = .1\textwidth, valign = c, clip, trim = 0 0 0 0]{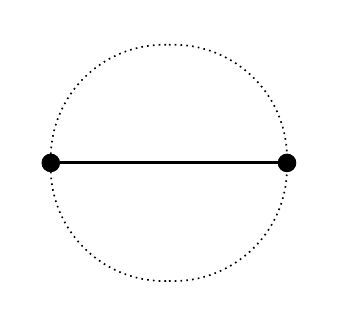} \right)
}_\text{N\textsuperscript{2}LO corrections}
\end{equation*}
\caption{\label{fig:lo nlo diags}%
Feynam diagrams that contribute to the effective potential at LO, NLO, and N\textsuperscript{2}LO.
All lines denote resummed exact propagators;
Wave lines correspond to gauge boson, dotted lines to ghost fields, dashed lines to unphysical Goldstone bosons, and straight lines to physical Higgs bosons.
}
\end{figure}

\Cref{fig:lo nlo diags} depicts all diagrams that contribute to $V^\prime_\text{eff}$ at LO, NLO, and N\textsuperscript{2}LO.
The one-loop diagrams involving gauge boson, ghost field, and Goldstone boson propagators contribute at leading order
and contain thermal corrections that contribute at NLO in the soft mass regime $\varphi \sim T$.
The evaluation of these thermal corrections is the main focus of this section.
The remaining diagrams contribute at N\textsuperscript{2}LO.
The one-loop diagram involving the physical Higgs boson propagator only contributes at N\textsuperscript{2}LO because its prefactor
\begin{align}
\partial_\varphi m_h^2 = 6 \, \lambda \, \varphi \sim g^4 \varphi
\end{align}
is suppressed by the small value of $\lambda$.
Likewise, the two-loop diagrams each contain two vertices that scale as $g^2 \varphi$,
so that the diagrams as a whole scale like $g^4 \varphi_\text{min}^2 T^2 \sim g^2 T^4$ in the hard hard mass regime
and like $g^4 \varphi_\text{soft}^2 T^2 \sim g^4 T^4$ in the soft mass regime.
Therefore, they are always suppressed by a factor of $g^2$ compared to the LO contributions.

To start, we consider the effective potential in the hard mass regime, where thermal resummation is unnecessary.
By inserting the tree-level inverse propagators and spectral functions  given in \cref{eq:tree invprop,eq:tree spec} into into the general formulae \cref{eq:inv prop var} and \cref{eq:1pi pot},
one finds that the three leading one-loop diagrams in \cref{fig:lo nlo diags} generate the contribution
\begin{align}\label{eq:1loop int}
\Delta V^\prime
&\supset V^\prime_\text{cw} + (\partial_\varphi m_A^2) \int \frac{\text{d}^4 k}{(2\pi)^4} 2 \Theta(k_0) f_B(k_0) \left[ - g^{\mu\nu} [\rho_{0 \, A}]_{\mu\nu} + \xi \left( \rho_{0 \, \chi} - 2 \rho_{0 \, c} \right) \right]
\\ \label{eq:1loop pot}
&= V^\prime_\text{cw} + (\partial_\varphi m_A^2) \left[ 3 I(m_A) + \xi ( I(m_\chi) - I(m_c) ) \right] \ ,
\end{align}
where \cite{Coleman:1973jx,Hirvonen:2021zej}
\begin{align}
\Delta V_\text{cw}^\prime (\varphi) &= \frac{3 \, g^4 \varphi^3}{16 \pi^2} \left( \ln \left( \frac{g^2 \varphi^2}{\Lambda^2} \right) - \frac13 \right)
\end{align}
is the zero-temperature Coleman-Weinberg contribution to the effective potential with $\Lambda$ as renormalization scale.
The function
\begin{align}\label{eq:therm int}
I(m) &= \frac12 \int \frac{\text{d}^3 \bm k}{(2\pi)^3} \frac{f_B(\omega_k)}{\omega_k} \ , &
\omega_k^2 &= \bm k^2 + m^2
\end{align}
is a standard thermodynamic integral.
Following the power counting established in \cref{subsec:pwr count}, we may neglect the difference between $m_\chi$ and $m_c$ at leading order.
The gauge-parameter dependent contributions in \cref{eq:1loop pot} then cancel each other,
and one obtains the derivative of the leading order potential
\begin{align}\label{eq:LO pot}
V^\prime_\text{LO} &=
- \mu^2 \varphi + \lambda \varphi^3 + \Delta V_\text{cw}^\prime (\varphi)
+ (\partial_\varphi m_A^2) \, 3 I(m_A) \ .
\end{align}
After integrating over $\varphi$, we thus recover the textbook result
\begin{align}
\Delta V_\text{LO} &= - \frac12 \mu^2 \varphi^2 + \frac14 \lambda \varphi^4 + V_\text{cw} + 3 J(m_A, 0) \ ,
\end{align}
where
\begin{align}
J(m_1, m_2) &= - T \int \frac{\text{d}^3 \bm k}{(2\pi)^3} \log \left( \frac{1 + f_B(\omega_1)}{1 + f_B(\omega_2)} \right) \ , &
\omega_i^2 &= \bm k^2 + m_i^2 \ .
\end{align}
Thermal resummation becomes relevant in the soft mass regime $\varphi \sim T$
because in-medium effects can then generate order one corrections to the dispersion relation of soft $\bm k \sim g T$ modes.
To see this, consider the behaviour of the thermodynamic integral in \cref{eq:therm int} for small masses $m \lesssim g T$.
Using a high-temperature expansion, one finds
\begin{align}\label{eq:high T exp}
I(m) &= \frac{T^2}{24} - \frac{m T}{8\pi} + \mathcal O(g^2 T^2) \ , &
I(m) - I(0) &\sim g T \ ,
\end{align}
where the non-analytic $\propto m T$ term is generated by a $\omega_k \to 0$ pole in the thermal integral \cref{eq:therm int}.
Inserting the expansion \eqref{eq:high T exp} into the LO potential \eqref{eq:LO pot} and comparing the result with the power counting in \cref{eq:lo pwr count,eq:nlo nnlo pwr},
one finds that the $\propto T^2$ term contributes at LO and that the $ \propto m T$ term is an NLO correction.
Since the $T^2$ term is dominated by hard $\bm k \sim T$ modes, it is not affected by thermal resummation.
In contrast, the $m T$ term is dominated by soft $\bm k \sim g T$ modes and thus sensitive to in-medium effects that alter this dispersion relation.
In other words, thermal resummation does not change the LO potential in \cref{eq:LO pot} but is crucial for obtaining the correct NLO contributions.

To compute the NLO potential, we now implement the resummation scheme defined in \cref{sec:2pi resum}
and replace the Goldstone boson, ghost field, and gauge boson spectral functions in \cref{eq:1loop pot}
with resummed spectral functions that are obtained by solving the relevant propagator equations of motion.
For any scalar field, including the Goldstone boson and ghost fields, the solution is given as \cite{Prokopec:2003pj,Berges:2015kfa}.
\begin{align}\label{eq:resummed scalar spectral}
\rho_s &= \frac{\Pi_s^{\mathcal A}}{\left( k^2 - m_s^2 - \Pi_s^{\mathcal H} \right)^2 + \Pi_s^{\mathcal A \, 2}} \ , &
s &= \chi, c \ ,
\end{align}
where 
$\Pi_s^{\mathcal H}$ and $\Pi_s^{\mathcal A}$ are the real and imaginary parts of the retarded self-energies
\begin{align}\label{eq:scalar self energies}
\ii \Pi_s^r &= \Pi_s^{++} - \Pi_s^{+-} = \ii \Pi_s^{\mathcal H} + \Pi_s^{\mathcal A}
\ , &
\Pi_s^{ab} (x,y) &= - \frac{ab}{c_s} \frac{\delta \Gamma_2}{\delta G_s^{ba} (y,x)}
\ .
\end{align}
The shape of the photon spectral function is more complicated because thermal corrections generate Lorentz structures that are not present at zero temperature,
and because previous results for thermal gauge-boson propagators are only valid in the symmetric phase \cite{Weldon:1996kb,Gorda:2023zwy}.
Following the approach used in \cite{Gorda:2023zwy}, we write the spectral function as
\begin{align}\label{eq:photon spec decomp}
[\rho_A]_{\mu\nu} &=
\mathds P^T_{\mu\nu} \rho_T + \mathds P^L_{\mu\nu} \rho_L + + \mathds P^C_{\mu\nu} \rho_C + \mathds P^D_{\mu\nu} \rho_D \ ,
\end{align}
where the projectors $\mathds P^T_{\mu\nu}$, $\mathds P^L_{\mu\nu}$, and $\mathds P^D_{\mu\nu}$ are defined like in \cref{eq:proj def}.
The final object
\begin{align}
\mathds P^C_{\mu\nu} &= \frac{n_\mu k_\nu + k_\mu n_\nu}{|\bm k|} \ , &
n_\mu &= g_{0\mu} - \frac{k_0 k_\mu}{k^2} \ ,
\end{align}
is technically not a projection operator because it is not idempotent but nevertheless necessary to capture all possible Lorentz structures.
The retarded photon self-energy
\begin{align}\label{eq:self en def}
\ii \Pi^r_{\mu\nu} &= \Pi^{++}_{\mu\nu} - \Pi^{+-}_{\mu\nu} = \ii \Pi^{\mathcal H}_{\mu\nu} + \Pi^{\mathcal A}_{\mu\nu} \ , &
\Pi^{ab}_{\mu\nu} (x,y) &= - 2 ab \frac{\delta \Gamma_2}{\delta G_{\nu\mu}^{ba} (y,x)}
\end{align}
is likewise decomposed into the components
\begin{align}\label{eq:self en decomp}
\Pi^r_{\mu\nu} &=
\mathds P^T_{\mu\nu} \Pi^r_T + \mathds P^L_{\mu\nu} \Pi^r_L + + \mathds P^C_{\mu\nu} \Pi^r_C + \mathds P^D_{\mu\nu} \Pi^r_D \ .
\end{align}
In \cref{app:photon spectral}, we use this decomposition to generalize the symmetric phase results \cite{Weldon:1996kb,Gorda:2023zwy}
by deriving an approximate expression for the photon spectral function that is also valid in the broken phase.
\\

To proceed further, one has to compute the the (anti-)hermitian self-energies defined by \cref{eq:scalar self energies,eq:self en def}.
In order to obtain the correct NLO potential, we have to include all $\mathcal O(g^2)$ contributions that
play a role in regulating the IR divergence associated with the $\omega_k \to 0$ limit in \cref{eq:therm int}.
To extract only the regulating contributions, we evaluate the self-energies at zero external momenta and set $\varphi \to 0$.
In this limit, the anti-hermitian self-energies $\Pi_s^{\mathcal A}$ vanish.
The hermitian self-eneriges are replaced by the usual thermal masses
\begin{align}
\delta m_s^2 &= \lim_{\varphi \to 0} \delta \bar m_s^2 \ , &
\delta \bar m_s^2 &= \lim_{k_0 \to 0} \lim_{\bm k \to 0} \Pi_s^{\mathcal H} \ ,
\end{align}
which are well-known in the literature.
For the photon, one has \cite{Farakos:1994kx,Hirvonen:2021zej}%
\footnote{Technically, the results in \cite{Farakos:1994kx,Hirvonen:2021zej} were obtained in the imaginary time formalism rather than the real time formalism,
but this turns out to be irrelevant in this case.}
\begin{align}
\delta m_L^2 &= m_D^2 &
&\text{and}&
\delta m_T^2 &= \delta m_C^2 = \delta m_D^2 = \mathcal O(g^4) \ , 
\end{align}
where
\begin{align}
m_D^2 &= \frac13 g^2 T^2 + \mathcal O(g^4)
\end{align}
is the square of the Debye mass.
For the $\chi$ and ghost fields, one likewise obtains \cite{Farakos:1994kx,Hirvonen:2021zej}
\begin{align}
\delta m_\chi^2 &= \frac14 g^2 T^2 + \mathcal O(g^4) \ , &
\delta m_c^2 &= \mathcal O(g^4) \ .
\end{align}
By inserting these thermal masses into \cref{eq:resummed scalar spectral,eq:resummed photon spectral} and setting the anti-hermitian self energies to zero, we thus find the resummed spectral functions
\begin{subequations}\label{eq:resummed spec}
\begin{align}
\rho_s &= \sign(k_0) \pi \delta(k^2 - M_s^2) \ , &
s &= c, \chi
\end{align}
and
\begin{align}
\rho_A^{\mu\nu} &=
- \mathds P_T^{\mu\nu} \sign(k_0) \pi \delta(k^2 - M_A^2) - \mathds P_L^{\mu\nu} \sign(k_0) \pi \delta(k^2 - m_A^2)  - \mathds P_D^{\mu\nu} \, \xi \, \rho_c
\ ,
\end{align}\end{subequations}
where
\begin{align}
M_A^2 &= m_A^2 + m_D^2 \ , &
M_\chi^2 &= - \mu^2 + m_c^2 + \delta m_\chi^2 \ , &
M_c^2 &= m_c^2 
\end{align}
are resummed mass parameters.
To obtain the NLO potential, we finally insert these spectral functions into the one-loop contribution \cref{eq:1loop int}
and subtract the unresummed expressions, which are already included in the LO potential.
This yields the derivative of the potential%
\footnote{In principle, \cref{eq:nlo vprime} only holds for $M_0^2 = - \mu^2 + m_D^2 > 0$,
which is not true for sufficiently small temperatures.
For the following discussion, we assume that temperature is large enough to keep
the symmeric phase minimum $\varphi = 0$ meta-stable, so that $\lim_{\varphi \to 0} V_\text{LO}^{\prime\prime} > 0$.
This condition implies $\mu < \nicefrac{g T}{\sqrt{12}} < m_D$.}
\begin{align}\label{eq:nlo vprime}
V_\text{NLO}^\prime &= (\partial_\varphi m_A^2) \left[ I (M_A) - I(m_A) \right] + (\partial_\varphi m_c^2) \left[ I(M_\chi) - I (m_c) \right] \ ,
\end{align}
and, after integrating over $\varphi$,
\begin{align}\label{eq:nlo pot}
\Delta V_\text{NLO} &= J(M_A, m_D) - J(m_A, 0) + J(M_\chi, M_0 ) - J(m_c, 0) \ , &
M_0^2 &\equiv M_\chi^2 - m_c^2 \ .
\end{align}
Hence, we obtain the full potential
\begin{align}\nonumber
\Delta V_\text{eff}
&\equiv \Delta V_\text{LO} + \Delta V_\text{NLO} \\ \label{eq:full potential}
&= - \frac12 \mu^2 \varphi^2 + \frac14 \lambda \varphi^4 + V_\text{cw} + 2 \, J(m_A, 0) + J(M_A, m_D) + J(M_\chi, M_0) - J(m_c, 0) \ .
\end{align}
In Landau gauge, where $\xi = 0$, this reduces to
\begin{align}\label{eq:landau potential}
\Delta V_\text{eff}
&\overset{\xi = 0}{=} - \frac12 \mu^2 \varphi^2 + \frac14 \lambda \varphi^4 + V_\text{cw} + 2 \, J(m_A, 0) + J(M_A, m_D) \ .
\end{align}
At a surface level, this result appears to be very similar to the Daisy resummed potentials given \eg in \cite{Arnold:1992rz,Wainwright:2011qy,Christiansen:2025xhv}.
However, there are two key differences between our procedure and the usual Daisy resummation scheme:
\begin{enumerate}
\item
In a Daisy resummation, all particles are usually treated on an equal footing, and the potential is computed by summing over all diagrams that contribute at a given loop order.
Here, we found that the physical Higgs boson loop only enters at N\textsuperscript{2}LO and thus should be treated separately.
Likewise, thermal corrections to the ghost field masses are suppressed compared to the thermal Goldstone and gauge boson masses and should not be included in the NLO potential.
Therefore, a naive Daisy resummation effectively mixes different orders in perturbation theory and ultimately yields an inconsistent result.
\item
Second, our procedure is derived from first principles of thermal field theory and therefore well-defined at any order in perturbation theory.
Starting at N\textsuperscript{2}LO, one has to account for the antihermitian self-energies $\Pi_s^{\mathcal A}$,
which generate a finite width for the resummed spectral functions.
It is then no longer possible to interpret their thermal corrections as simple mass shifts.
\end{enumerate}
Nevertheless, we find that our NLO result can be recovered from the Daisy resummed effective potential in \cite{Wainwright:2011qy}
by shifting the potential to ensure that $\Delta V_\text{eff}(0) = 0$ and dropping all terms that formally contribute at N\textsuperscript{2}LO or higher.
In \cref{app:nielsen}, we also show explicitly that our potential satisfied the Nielsen identity need for gauge-parameter independent predictions of the bubble nucleation rate.
This suggests that the Daisy resummation can indeed yield gauge independent predictions for phase transition observables,
provided that one imposes a consistent power counting.%
\footnote{We appreciate that this is finding is consistent with the message advocated by \cite{Lofgren:2023sep}.
}

\subsection{Comparison with dimensional reduction}
\label{subsec:comparison}

As a final cross-check, we test that the full potential \eqref{eq:full potential} is consistent with previous results from dimensional reduction.
In the soft mass regime $\varphi \sim T$, where dimensional reduction is a viable method for computing the effective potential,
we may use the high-temperature expansion
\begin{align}
J(m_1, m_2) &= \frac{T^2}{24} (m_1^2 - m_2^2) - \frac{T}{12\pi} ( m_1^3 - m_2^3) + \mathcal O(g^4) \ .
\end{align}
After inserting this expansion into the full potential \eqref{eq:full potential} and reordering
the individual contributions to account for the fact that the $T m^3$ terms are suppressed compared to the leading $T^2 m^2$ terms,
this yields
\begin{align}\label{eq:small mass pot}
\Delta V_\text{LO} &\overset{\varphi \sim T}{=} \frac12 M_0^2 \varphi^2 \ , &
\Delta V_\text{NLO} &\overset{\varphi \sim T}{=} - \frac{T}{12\pi} \left( 2 m_A^3 + M_A^3 - m_D^3 + M_\chi^3 - M_0^3 - m_c^3 \right) \ ,
\end{align}
where
\begin{align}
M_0^2 &= M_\chi^2 - m_c^2 \sim g^2 T^2
\end{align}
has to scale like the photon mass $g^2 T^2$ to ensure
that all contributions to the leading potential are comparable in the vicinity of the broken phase minimum $\varphi_\text{min} \sim \nicefrac{T}{g}$.
We now compare this expression with the effective theory result given in equations (3.22) and (3.23) of \cite{Hirvonen:2021zej},
which also computes the Abelian Higgs effective potential a general covariant gauge.
That computation is consistent in the regime
\begin{align}
\lambda &\sim g^3 \ , &
\mu^2 &\sim g^2 T^2 \ , &
M_0^2 &\sim g^3 T^2 \ , &
\varphi^2 &\lesssim \nicefrac{T^2}{g} \ .
\end{align}
This power counting differs from our counting, where $\lambda \sim g^4$ is smaller while $M_0^2 \sim g^2 T^2$ is taken to be larger.
To meaningfully compare our result with \cite{Hirvonen:2021zej}, we have to impose the same power counting for both computations.
In the remainer of this section, we thus count
\begin{align}
\lambda &\sim g^4 \ , & 
M_0^2 &\sim g^3 T^2 \ .
\end{align}
With this power counting, both contributions in \cref{eq:small mass pot} are the same size.
Dropping the Goldstone boson and ghost field terms, which are now formally of higher order, one finds
\begin{align}\label{eq:alt pwr count}
\Delta V_\text{LO} &\overset{\varphi \sim T}{=} \frac12 M_0^2 \varphi^2 - \frac{T}{12\pi} \left( 2 m_A^3 + M_A^3 - m_D^3 \right) &
\text{for}&&
M_0^2 &\sim g^3 T^2 \ .
\end{align}
Up to a negligible term that depends on $\lambda$ and an overall field-independent shift,
this is identical to the result reported in equation (3.22) of \cite{Hirvonen:2021zej}.
The shift is present because we compute the difference between the potential in the broken and symmetric phase.
Hence, the potential is normalized such that
\begin{align}
\lim_{\varphi \to 0} \Delta V_\text{eff} = 0 \ .
\end{align}
In contrast, the expression in (3.22) of \cite{Hirvonen:2021zej} yields the limit
\begin{align}
\lim_{\varphi \to 0} V_\text{eff} = T \lim_{\varphi \to 0} V^\text{3d}_\text{eff,LO} = \frac{T}{12\pi} m_D^3 \ ,
\end{align}
so that the potential difference is really given as
\begin{align}
\Delta V_\text{eff} &= T \left( V^\text{3d}_\text{eff,LO} - \frac{T m_D^3}{12\pi} \right) = \frac12 M_0^2 \varphi^2 - \frac{T}{12\pi} \left( 2 m_A^3 + M_A^3 - m_D^3 \right)  + \text{$\lambda$-dependent terms} \ .
\end{align}
This is the same expression as in \cref{eq:alt pwr count}.
Hence, the full potential in \cref{eq:full potential} is indeed consistent with dimensional reduction for small values of the condensate $\varphi$.

\section{Conclusions and outlook}
\label{sec:conclusion}

In this work, we have shown that 2PI effective action techniques can be used to
consistently resum effective potentials in cases that go beyond the regime of applicability of dimensional reduction.
To this end, we developed a novel resummation scheme that uses
a real-time approach and to compute a resummed expression for the potential derivative $V_\text{eff}^\prime$.

We used this scheme to compute the next-to-leading order
effective potential \eqref{eq:full potential} for a strong transition of the Abelian Higgs model in a general covariant gauge,
and have shown explicitly that it satisfies the leading-order Nielsen identity
needed for gauge-parameter independent predictions of the bubble nucleation rate.
Our computation thus improves the old result in \cite{Amelino-Camelia:1993rvt} by extending it to very strong transitions and general covariant gauges.
The final potential is consistent with existing results from dimensional reduction
and can be recovered from the Daisy resummed one-loop effective potential
by dropping all terms that contribute N\textsuperscript{2}LO or
higher according to the power counting defined in \cref{subsec:pwr count}.
Critically, this power counting differs from the counting that
usually governs thermal phase transitions in Higgs gauge theories.
In particular, the fact that dimensional reduction is not applicable
in the broken phase implies that the Higgs self-interaction must be very small and scale like $\lambda \sim g^4$,
while the effective Higgs mass parameter $M_0^2 = \frac14 g^2 T^2 - \mu^2$
 is expected to be larger than in weaker first-order transitions.
\\

There are several promising directions for further investigations.
First of all, it would be interesting to apply the procedure developped in this work to strong transitions in more complex Higgs-gauge theories,
including models with Fermions and non-Abelian gauge interactions, and ultimately to realistic \SM extensions.
Established resummation schemes for ab-initio computations of the effective potential,
including dimensional reduction, are expected to break down for very strong first-order phase transitions.
In contrast, the scheme presented is viable for arbitrarily strong transitions,
and thus makes it possible to derive robust predictions for \SGWBs in cases that established techniques cannot handle.
As a first test to see how sensitive \SGWBs from very strong transitions are to thermal resummation,
it would be interesting to use the full potential given in \cref{eq:full potential}
to predict the \SGWB for a strong transition in the Abelian Higgs model.
Furthermore, it would be interesting to compare our results with other resummation methods that attempt to go beyond dimensional reduction,
including in particular the heat-kernel method \cite{Bandyopadhyay:2026nrv}.
Finally, it would be useful to achieve a full N\textsuperscript{2}LO computation of the effective potential,
as this would significantly improve the ability to estimate theoretical uncertainties associated \eg with renormalization scale dependence of the effective potential.
 
\section*{Acknowledgments}

We gratefully thank Philipp Schicho for insightful comments and valuable feedback on the content of this manuscript.
The author was supported by the Swiss National Science Foundation 
(SNSF) under grant Nr. P500PT-217885.

%=======================================================================================
%Backmatter
\appendix
\appendixpage
\addappheadtotoc
\crefalias{section}{appendix}

\section{Nielsen identity}
\label{app:nielsen}

The conventional zero-temperature 1PI effective action obeys the Nielsen identity \cite{Metaxas:1995ab}
\begin{align}\label{eq:zero T Nielsen}
\xi \partial_\xi \Gamma_\text{1PI}
&= \frac{\ii \, g}2 \int \vspace{-3pt} \text{d}^4 x \, \text{d}^4 y \ \frac{\delta \Gamma_\text{1PI}}{\delta \varphi (x)} \left\langle \mathcal T \left\{ \chi(x) \, \overline c(x) \, c(y) \, F(y) \right\} \right\rangle_\text{1PI} \ ,
\end{align}
where $\mathcal T$ is the usual time-path ordering operator and $F$ is the gauge fixing functional defined in \cref{eq:gauge fix}.
To obtain the corresponding identity for the closed-time path effective action defined in \cref{eq:1pi ctp action},
one has to replace the ordinary time-integrals in \cref{eq:zero T Nielsen} with integrals over the closed time path.
This amounts to the substitution
\begin{align}
\int \text{d}^4 x \to \sum_a a \int \text{d}^4 x \ ,
\end{align}
where each field that depends on the coordinate $x$ also picks up a corresponding time-path index $a$.
Hence, the real-time effective action obeys the Nielsen identity
\begin{subequations}\begin{align}
\xi \partial_\xi \Gamma_\text{1PI}
&= \frac{\ii}2 g \sum_{ab} ab \int \vspace{-3pt} \text{d}^4 x \, \text{d}^4 y \, \frac{\delta \Gamma_\text{1PI}}{\delta \varphi^a(x)} \left\langle \mathcal T_c \left\{ \chi^a (x) \, \overline c^a(x) \, c^b(y) \, F^b(y) \right\} \right\rangle_\text{1PI}
\ .
\end{align}\end{subequations}
Using the definition of the effective potential interms of the 1PI effective action, this yields the constraint
\begin{align}\label{eq:full nielsen}
\xi \partial_\xi V_\text{eff} &= C \, \partial_\varphi V_\text{eff} \ ,
\end{align}
where
\begin{subequations}\begin{align}
C &\equiv - \frac{\ii}4 g \sum_{ab} b \int \vspace{-3pt} \text{d}^4 y \, \left\langle \mathcal T_c \left\{ \chi^a (0) \, \overline c^a(0) \, c^b(y) \, F^b(y) \right\} \right\rangle_\text{1PI} \\
\label{eq:nielsen coeff}
&= \frac{m_c^2}{2 \varphi} \int \vspace{-3pt} \frac{\text{d}^4 p}{(2\pi)^4} \, \left( 1 + 2 f_B \right)
\left[ G_\chi^{\mathcal H} (p) \rho_c (p) + \rho_\chi (p) G_c^{\mathcal H} (p) \right] \left( 1 + \mathcal O(g) \right)
\end{align}\end{subequations}
is the Nielsen coefficient.
By expanding \cref{eq:full nielsen} to its first non-vanishing order, and using that the leading-order effective potential is manifestly gauge-parameter independent,
one obtains the leading-order Nielsen identity
\begin{align}
\xi \partial_\xi V_\text{NLO} &= C_\text{LO} V^\prime_\text{LO} \ ,
\end{align}
where $C_\text{LO}$ is now the leading order Nielson coefficient.
In the remainder of this section, we show that the leading and next-to-leading potentials given in \cref{eq:LO pot,eq:nlo pot} satisfy this identity.

To obtain a well-defined result for $C_\text{LO}$, one has to keep track of the difference between the masses of the unphysical Goldstone boson and the ghost field in \cref{eq:nielsen coeff}.
Since the square mass difference $m_\chi^2 - m_c^2 \sim g^2 T^2$ is perturbatively small, thermal corrections to both masses can enter into the leading order coefficient $C_\text{LO}$.
To compute the Nielsen coefficient for arbitrary $\varphi$,
we need expressions for the thermal mass corrections with a finite condensate $\varphi$.
After a straightforward computation, one finds
\begin{subequations}\begin{align}\label{eq:mass pot rel}
\delta \bar m_\chi^2
&= - m_\chi^2 + m_c^2 + \frac{V_\text{LO}^\prime (\varphi)}{\varphi} \ , &
\delta \bar m_c^2
&= \frac{\partial_\varphi m_c^2}{\varphi} \left[ I(m_c) - I(0) \right] + \text{vacuum terms} \ .
\end{align}\end{subequations}
After inserting the explicit shape of the spectral functions and replacing
\begin{align}
m_s^2 \to \bar M_s^2 &\equiv m_s^2 + \delta \bar m_s^2 \ , &
s &= c, \chi \ ,
\end{align}
the integral in \cref{eq:nielsen coeff} becomes
\begin{subequations}\begin{align}
C
&= \frac1{2 \varphi} \frac{m_c^2}{\bar M_\chi^2 - \bar M_c^2} \int \vspace{-3pt} \frac{\text{d}^4 p}{(2\pi)^4} \, \left( 1 + 2 f_B \right) \sign(p_0) \pi 
\left[ \delta(p^2 - \bar M_\chi^2) - \delta(p^2 - \bar M_c^2)  \right] \left( 1 + \mathcal O (g) \right)
\\\label{eq:nielsen intermediate}
&= \frac{m_c^2}{V_\text{LO}^\prime + \varphi \, \delta \bar m_c^2} \left[ I(\bar M_\chi) - I(\bar M_c) \right] + \text{higher order} \ .
\end{align}\end{subequations}
The ghost field mass correction $\delta m_c^2$ does not enter into the leading order coefficient $C_\text{LO}$ because it is perturbatively suppressed compared to both $m_c^2$ and $\nicefrac{V_\text{LO}^\prime}{\varphi}$.
According to the power counting established in \cref{eq:nlo nnlo pwr}, the difference
\begin{align}
I( \bar M_\chi) - I(M_\chi)
\end{align}
is also a N\textsuperscript{2}LO correction to the effective potential, so that we should replace $I(\bar M_\chi) \to I(M_\chi)$ in \cref{eq:nielsen intermediate}.
Hence, we obtain our final result
\begin{align}
C_\text{LO} \, V_\text{LO}^\prime
&= m_c^2 \left[ I(M_\chi) - I(m_c) \right] \ .
\end{align}
Finally, we consider the gauge parameter variation of the next-to-leading order effective potential.
By inserting the explicit expression given in \cref{eq:nlo pot}, we find
\begin{align}
\xi \partial_\xi V_\text{NLO} &= m_c^2 \left[ I(M_\chi) - I(m_c) \right] \ .
\end{align}
Thus, we have shown explicitly that our result obeys the leading order Nielsen identity.

\section{Photon spectral function}
\label{app:photon spectral}

In this appendix, we derive an approximate expression for the resummed photon spectral function in the broken phase with $\varphi \neq 0$. 
To do so, we follow the same overall approach used in \cite{Weldon:1996kb,Gorda:2023zwy} to obtain finite temperature gauge boson propagators in the symmetric phase,
and extend it to include finite gauge boson masses.
By varying the 2PI effective action, one finds that the advanced and retarded photon propagators obey the equation of motion 
\begin{align}\label{eq:prop eom}
g_{\mu\nu} &=
\left( \left( 1 - \frac1{\xi} \right) k_\mu k^\lambda - (k^2 - m_A^2) \delta_\mu^{\ \lambda} + \Pi^{a,r \ \lambda}_{\, \mu} (k) \right) [G_A]^{a,r}_{\lambda\nu} (k) \ ,
\end{align}
where the advanced and retarded self energies are defined as in \cref{eq:self en def}.
To solve this equation, we define the projectors
\begin{align}
\mathds P^T_{\mu\nu} &= \delta_\mu^{\ i} \delta_\nu^{\ j} \left( g_{ij} + g_{00} \frac{k_i k_j}{\bm k^2} \right) \ , &
\mathds P^L_{\mu\nu} &= \frac{n_\mu n_\nu}{n^2} \ , &
\mathds P^C_{\mu\nu} &= \frac{n_\mu k_\nu + k_\mu n_\nu}{|\bm k|} \ , &
\mathds P^D_{\mu\nu} &= \frac{k_\mu k_\nu}{k^2} \ ,
\end{align}
where
\begin{align}
n_\mu &= g_{0\mu} - \frac{k_0 k_\mu}{k^2} \ , &
n^2 &= \frac{- \bm k^2}{k^2}
\end{align}
is an independent vector that is orthogonal to $k_\mu$.
The projectors $\mathds P^T$, $\mathds P^L$ and $\mathds P^D$ are mutually orthogonal, idempotent, and obey the sum rules given in \cref{eq:proj sum rules}.
The final projector $\mathds P^C$ is neither idempotent nor orthogonal to $\mathds P^D$ and $\mathds P^L$. 
Instead, it satisfies the relations
\begin{align}
\mathds P^{T \, \lambda}_{\, \mu} \mathds P^C_{\lambda\nu} &= \mathds P^{C \, \lambda}_{\, \mu} \mathds P^T_{\lambda\nu} = 0 \ , &
\mathds P^C_{\mu\lambda} \mathds P^{C \lambda}_{\ \nu} &= - \mathds P^L_{\mu\nu} - \mathds P^D_{\mu\nu} \ , \\
\mathds P^{L \, \lambda}_{\, \mu} \mathds P^C_{\lambda\nu} &= \mathds P^{C \, \lambda}_{\, \mu} \mathds P^D_{\lambda\nu} = \frac{n_\mu k_\nu}{|\bm k|} \ , &
\mathds P^{D \, \lambda}_{\, \mu} \mathds P^C_{\lambda\nu} &= \mathds P^{C \, \lambda}_{\, \mu} \mathds P^L_{\lambda\nu} = \frac{k_\mu n_\nu}{|\bm k|} \ .
\end{align}
Using the above projectors, we decompose the propagator and the self-energy as
\begin{subequations}\begin{align}
\Pi^{a,r}_{\mu\nu} &= \mathds P^T_{\mu\nu} \Pi^{a,r}_T + \mathds P^L_{\mu\nu} \Pi^{a,r}_L + \mathds P^C_{\mu\nu} \Pi^{a,r}_C + \mathds P^D_{\mu\nu} \Pi^{a,r}_D \ , \\
[G_A]^{a,r}_{\mu\nu} &= \mathds P^T_{\mu\nu} [G_A]^{a,r}_T + \mathds P^L_{\mu\nu} [G_A]^{a,r}_L + \mathds P^C_{\mu\nu} [G_A]^{a,r}_C + \mathds P^D_{\mu\nu} [G_A]^{a,r}_D \ .
\end{align}\end{subequations}
Inserting these components into the propagator equation of motion \cref{eq:prop eom} yields  the five equations
\begin{subequations}\begin{align}
1 &= - (k^2 - m_A^2 - \Pi^{a,r}_T) [G_A]^{a,r}_T \ , \\
1 &= - (k^2 - m_A^2 - \Pi^{a,r}_L) [G_A]^{a,r}_L - \Pi^{a,r}_C [G_A]^{a,r}_C \ , \\
1 &= - \left( \frac{k^2 - m_c^2}{\xi} - \Pi^{a,r}_D \right) [G_A]^{a,r}_D - \Pi^{a,r}_C [G_A]^{a,r}_C \ , \\
0 &= - \left( \frac{k^2 - m_c^2}{\xi} - \Pi^{a,r}_D \right) [G_A]^{a,r}_C + \Pi^{a,r}_C [G_A]^{a,r}_L \ , \\
0 &= - (k^2 - m_A^2 - \Pi^{a,r}_L) [G_A]^{a,r}_C + \Pi^{a,r}_C [G_A]^{a,r}_D \ .
\end{align}\end{subequations}
They have the solution
\begin{subequations}\begin{align}
[G_A]^{a,r}_T &= \frac{-1}{k^2 - m_A^2 - \Pi^{a,r}_T} \ , \\
[G_A]^{a,r}_L &= \frac{-1}{k^2 - m_A^2 - \Pi^{a,r}_L + \xi \frac{(\Pi^{a,r}_C)^2}{k^2 - m_c^2 - \xi \Pi^{a,r}_D} } \ , \\
[G_A]^{a,r}_C &= \frac{-1}{k^2 - m_A^2 - \Pi^{a,r}_L} \frac{\xi \Pi^{a,r}_C}{k^2 - m_c^2 - \xi \Pi^{a,r}_D + \xi \frac{(\Pi^{a,r}_C)^2}{k^2 - m_A^2 - \Pi^{a,r}_L} } \ , \\
\label{eq:D prop}
[G_A]^{a,r}_D &= \frac{-\xi}{k^2 - m_c^2 - \xi \Pi^{a,r}_D + \xi \frac{(\Pi^{a,r}_C)^2}{k^2 - m_A^2 - \Pi^{a,r}_L} } \ .
\end{align}\end{subequations}
In accordance with the discussion in section \cref{subsec:eff pot}, we evaluate the self-energies for zero external momenta and set $\varphi \to 0$.
We can then use the properties of the advanced and retarded propagators in the symmetric phase to constrain the shape of the self-energies.
In particular, they have to obey the consistency relation \cite{Weldon:1996kb,Gorda:2023zwy} 
\begin{align}
\Pi^{a,r}_D = \frac{(\Pi^{a,r}_C)^2}{k^2 - \Pi^{a,r}_L} \ ,
\end{align}
which implies that $\Pi^{a,r}_D$ is negligible at $\mathcal O(g^2)$.
Also neglecting terms involving $\Pi^{a,r}_C$ if they do not contribute at $\mathcal O(g^2)$, the advanced and retarded propagators become
\begin{align}\label{eq:D prop}
[G_A]^{a,r}_{T,L} &= \frac{-1}{k^2 - m_A^2 - \Pi^{a,r}_{T,L}} \ , &
[G_A]^{a,r}_C &= 0 \ , &
[G_A]^{a,r}_D &= \frac{-\xi}{k^2 - m_c^2} \ .
\end{align}
For the photon spectral function, this yields our final result
\begin{align}\label{eq:resummed photon spectral}
[\rho_A]_{\mu\nu} &= \frac1{2\ii} \left( [G_A]^a_{\mu\nu} - [G_A]^r_{\mu\nu} \right) 
= \mathds P^T_{\mu\nu} \rho_T + \mathds P^L_{\mu\nu} \rho_L + \mathds P^D_{\mu\nu} \rho_D \ ,
\end{align}
where
\begin{align}
\rho_{\, T,L} &= \frac{- \Pi^{\mathcal A}_{T,L}}{(k^2 - m_A^2 - \Pi_{T,L}^{\mathcal H})^2 + \Pi_{T,L}^{\mathcal A 2}} \ , &
\rho_D &= - \xi \pi \sign(k_0) \delta(k^2 - m_c^2) \ .
\end{align}

\printbibliography

\end{document}